\documentclass[noshowpacs,prc,preprint,nofootinbib,showkeys,onecolumn]{revtex4}
\pdfoutput=1

\usepackage{amssymb}
\usepackage{graphicx}
\usepackage{amsmath}
\usepackage{ulem}
\usepackage{subfigure}
\usepackage{bbold}
\usepackage{yfonts} 
\usepackage{placeins}
\usepackage{bm} 
\usepackage{nicefrac}
\usepackage{slashed} 
\usepackage{color}
\usepackage[dvipsnames]{xcolor}
\usepackage{hyperref}
\hypersetup{colorlinks=true, citecolor=Red, linkcolor=Blue, urlcolor=CornflowerBlue, linktocpage=true}
\usepackage[makeroom]{cancel}
\newcommand{\bea}{\begin{eqnarray}}
\newcommand{\eea}{\end{eqnarray}}

\def\be{\begin{equation}}
\def\ee{\end{equation}}
\newcommand{\bel}[1]{\begin{eqnarray}\label{#1}}
\newcommand{\eel}{\end{eqnarray}}
\def\barr{\begin{array}}
\def\earr{\end{array}}
\def\beq{\begin{eqnarray}}
\def\eeq{\end{eqnarray}}
\def\bfig{\begin{figure}}
\def\efig{\end{figure}}

\def\PsiRP{{\Psi_{\rm RP}}}
\def\PsiEP{{\Psi_{\rm EP}^{(1)}}}
\def\DPsi{{\Delta\Psi}}

\newcommand{\av}{{\boldsymbol a}} 
\newcommand{\bv}{{\boldsymbol b}} 
 
\newcommand{\nv}{{\boldsymbol n}} 

\newcommand{\vv}{{\boldsymbol v}}
\newcommand{\pv}{{\boldsymbol p}}

\newcommand{\xvhat}{\hat{\boldsymbol x}}

\newcommand{\yvhat}{\hat{\boldsymbol y}}

\newcommand{\zvhat}{\hat{\boldsymbol z}}

\newcommand{\pvl}{{\boldsymbol p}_\Lambda}
\newcommand{\pvhat}{\hat{{\boldsymbol p}}}
\newcommand{\Lv}{{\boldsymbol L}}
\newcommand{\Lvhat}{\hat{\boldsymbol L}}
\newcommand{\Kv}{{\boldsymbol K}}
\newcommand{\Kvhat}{\hat{\boldsymbol K}}
\newcommand{\Pv}{{\boldsymbol P}}
\newcommand{\lamv}{{\boldsymbol \lambda}}
 
\begin{document}
%%%%%%%%%%%%%%%%%%%%%%%%%%%%%%%%%%%%%%%%%%
 
\title{On the interpretation of $\Lambda$ spin polarization measurements}  

	\author{Wojciech Florkowski}
	\email{wojciech.florkowski@uj.edu.pl}
	 \affiliation{Institute of Theoretical Physics, Jagiellonian University,  PL-30-348 Krak\'ow, Poland}
	  
    \author{Radoslaw Ryblewski} 
    \email{radoslaw.ryblewski@ifj.edu.pl} 
    \affiliation{Institute of Nuclear Physics Polish Academy of Sciences, PL-31342 Krakow, Poland}

\date{\today} 
\bigskip 
  
\begin{abstract}
The physics interpretation of the recent measurements of the spin polarization of $\Lambda$~hyperons produced in relativistic heavy-ion collisions is discussed. We suggest that the polarization measured in the $\Lambda$ rest frame should be projected along the direction of the total angular momentum that is first transformed to the same frame, and only then averaged over $\Lambda$'s with different momenta in the center-of-mass frame. While this procedure does not affect the current measurements done in a broad transverse-momentum range, it may become important (represent a correction of about 10\%) for the most energetic hyperons under study (with transverse momenta reaching 4--5 GeV/c). The proposed treatment is  generally more appropriate for relativistic $\Lambda$'s. Throughout the paper, we deliver explicit expressions for various boosts, rotations, and transformations of angular distributions, which may help to compare model predictions with the experimental results. 
\end{abstract}
    
\keywords{spin polarization, Lambda hyperons, relativistic heavy-ion collisions, proton-proton collisions}
	 
\maketitle
%%%%%%%%%%%%%%%%%%%%%%%%%%%%%%%%%%%%%%%%%%%%
\section{Introduction}
%%%%%%%%%%%%%%%%%%%%%%%%%%%%%%%%%%%%%%%%%%%%

For a few decades now the phenomenon of spin polarization of the $\Lambda$ hyperons produced in proton-proton and heavy-ion collisions has been an intriguing topic of both experimental and theoretical investigations \cite{Bunce:1976yb,BOURRELY198095,Panagiotou:1989sv,Becattinibook}.
For example, the longitudinal polarization of the ${\bar \Lambda}$ hyperons was discussed in 1980s as a possible signal of the quark-gluon plasma formation~\cite{Jacob:1987sj}. However,  the first heavy-ion experiments that measured the $\Lambda$ spin polarization in Dubna~\cite{Anikina:1984cu} and at CERN~\cite{Bartke:1990cn} reported negative results. More recently, several theoretical predictions of the global spin polarization signal in A+A collisions were given in Refs.~\cite{Liang:2004ph,Betz:2007kg,Voloshin:2004ha}. These works predicted a rather substantial experimental signal, of the order of 10\%, and were not confirmed by the STAR data of 2007 \cite{Abelev:2007zk}. The idea of a non-vanishing global polarization reappeared in the context of statistical physics and equilibration of spin degrees of freedom~\cite{Becattini:2007nd,Becattini:2007sr,Becattini:2013fla,Becattini:2013vja,Becattini:2016gvu}. The much smaller predictions of this approach \cite{Karpenko:2016jyx,Li:2017slc,Xie:2017upb,Sun:2017xhx} have been eventually observed by STAR~\cite{STAR:2017ckg,Adam:2018ivw} and independently by ALICE~\cite{Acharya:2019vpe}. This has triggered a vast theoretical interest that includes several highly debated topics: the importance of the spin-orbit coupling~\cite{Gao:2007bc,Chen:2008wh}, global equilibrium with a rigid rotation~\cite{Becattini:2009wh,Becattini:2012tc,Becattini:2015nva,Hayata:2015lga}, hydrodynamic~\cite{Florkowski:2017ruc,Florkowski:2017dyn,Florkowski:2019voj,Li:2020eon,Hu:2021lnx} and kinetic~\cite{Gao:2012ix,Chen:2012ca,Fang:2016vpj,Fang:2016uds,Florkowski:2018ahw,Weickgenannt:2019dks,Weickgenannt:2020aaf,Bhadury:2020puc,Bhadury:2020cop,Tinti:2020gyh} models of spin dynamics, anomalous hydrodynamics~\cite{Son:2009tf,Kharzeev:2010gr}, the Lagrangian formulation of hydrodynamics~\cite{Montenegro:2017rbu,Montenegro:2017lvf}, and hydrodynamic treatment of spin currents in the presence of torsion \cite{Gallegos:2021bzp}. For recent reviews of the experimental and theoretical situation see, for example, Refs.~\cite{Huang:2020xyr,Becattini:2020ngo,Florkowski:2018fap,Speranza:2020ilk,Bhadury:2021oat}.

\begin{figure}[t]
\begin{center}
\includegraphics[width=0.6\textwidth]{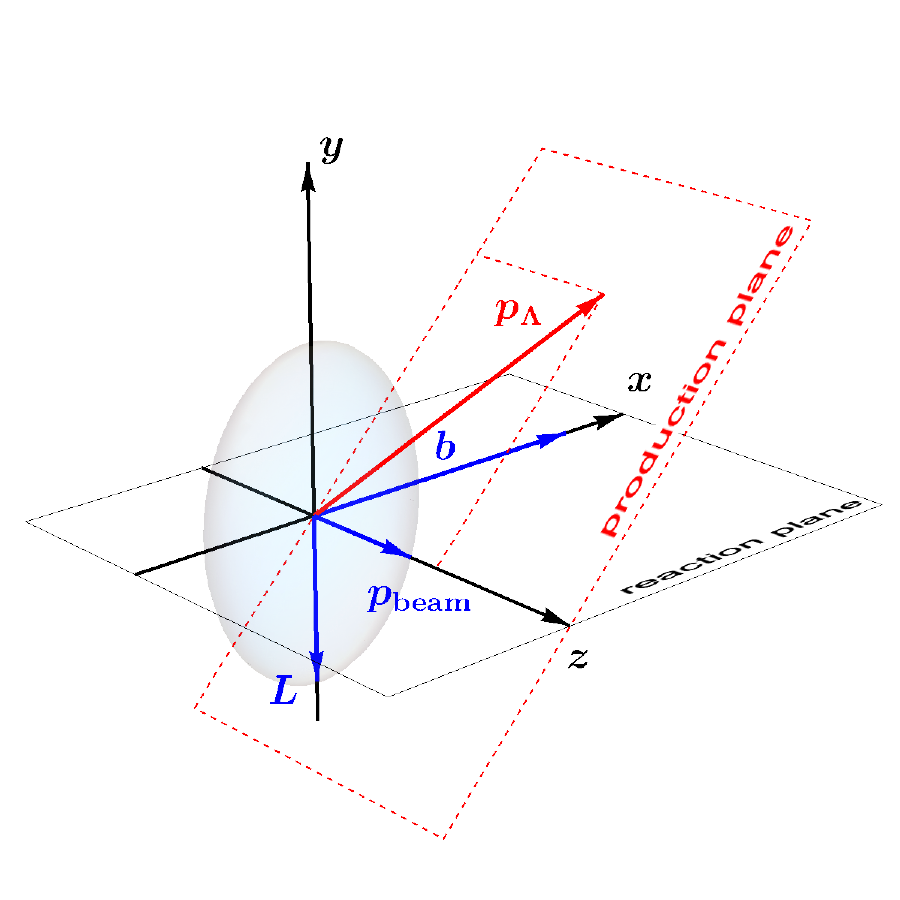}
\end{center}
\caption{The center-of-mass (COM) frame for non-central heavy-ion collisions. In this case both the reaction and production planes can be defined.}
\label{fig:COMhi}
\end{figure}

As the outcome of the spin polarization experiments, one commonly cites the magnitude of the polarization along a specific direction in the center-of-mass frame (COM). Most preferably,  the results refer to the direction that is orthogonal either to the reaction plane (RP, in non-central heavy-ion collisions) \cite{Becattinibook} or to the production plane (in proton-proton collisions) \cite{Bunce:1976yb,Panagiotou:1989sv}. In the case of heavy ions, the direction transverse to the reaction plane agrees with the direction of the total angular momentum of the system $\Lv$ (with the orientation of $\Lv$ opposite to the $y$ axis, see Fig.~\ref{fig:COMhi}). 

To determine the magnitude of the polarization in different directions, however, one first studies distributions of various three-momentum components of protons emitted in the weak decay $\Lambda \to p+\pi^-$, which are measured in the $\Lambda$ rest frame. As the COM frame and the $\Lambda$ rest frame are connected by the Lorentz transformation depending on the three-momentum of $\Lambda$, the spatial directions in these two frames are linked by a non-trivial relation. Consequently, relating the results obtained in the $\Lambda$ rest frame to the global angular momentum direction of the system requires that an appropriate Lorentz transformation is done before one describes such results in terms of the COM variables.

The STAR measurements~\cite{STAR:2017ckg,Adam:2018ivw,Acharya:2019vpe} indicate that the proton distributions in the $\Lambda$ rest frame are not isotropic and, consequently, unambiguously lead to the conclusion about the non-zero $\Lambda$ spin polarization. In our opinion, however, the interpretation of those results in the context of specific correlations between the spin direction of $\Lambda$'s and various directions in COM (in particular, the direction of the total orbital angular momentum $\Lv$) requires further clarifications because of at least two reasons. First, typically only one component of the polarization vector is measured --- the $y$-component in the $\Lambda$ rest frame. Second, the $y$ direction in the $\Lambda$ rest frame is different from the $y$ direction in COM.~\footnote{This effect has been neglected in the experimental analyses of the spin polarization of $\Lambda$'s, with a non-relativistic assumption that these two directions are the same. } Consequently, a complete understanding of the relation between the $\Lambda$ spin direction and the direction of the total orbital angular momentum in COM calls for a more detailed study of the effect connected with the boost to the $\Lambda$ rest frame. 

This is especially important if one interprets the result of the $\Lambda$ polarization measurements as an analog of the Einstein-de Haas or the Barnett effect \cite{dehaas:1915,Barnett:1935}. In this case, we suggest first to measure the projection of the spin polarization along the orbital angular momentum direction that is ``seen'' by a $\Lambda$ in its rest frame, and only then to make averaging over $\Lambda$'s with different momenta in COM. Such a method guarantees that the same physical direction is used for all $\Lambda$'s. We do not expect that such a procedure may change any qualitative conclusions about the global spin polarization but, in our opinion, it is more appropriate to establish the right magnitude of the polarization and its energy dependence. 

We note that an alternative method for measurements of the global polarization of $\Lambda$'s has been proposed in Ref.~\cite{Siddique:2017ddr}, where one demonstrates that the measurements can use quantities defined in the laboratory frame (instead of the quantities defined in the $\Lambda$'s rest frame). However, this work does not discuss the effects connected with a change of the orbital angular momentum direction due to the boosts, which is the main topic of the present analysis.

In this work, we give several explicit expressions for boosts, rotations, and transformations of angular distributions that can be useful whenever model predictions are compared with the experimental results. In particular, we give an expression for the form of the angular momentum in the $\Lambda$ rest frame that can be used to consistently project the polarization of $\Lambda$'s  measured in their rest frames. 

The paper is organized as follows: In the next section, we define the center-of-mass (COM) frame for heavy-ion and proton-proton collisions. In Sec. \ref{sec:Sprime} we introduce the canonical boost from COM to the $\Lambda$ rest frame and introduce the transformation of the total angular three-momentum from COM to the $\Lambda$ rest frame. Yet another $\Lambda$ rest frame, where the $\Lambda$ polarization is aligned with the $z$ axis, is introduced in Sec.~\ref{sec:Sstar}. The weak decay law for the process $\Lambda \to p + \pi^-$ is introduced in Sec.~\ref{sec:weakdecay}. Finally, in Sec.~\ref{sec:master} we discuss our main point regarding the projection of the measured polarization on the total angular momentum in COM. We summarize and conclude in Sec.~\ref{sec:conclusions}.   Several useful properties of the canonical boost and transformations of the angular distributions of protons are discussed in the two appendices.

\bigskip
{\it Conventions and notation:} Throughout the paper we use natural units with $\hbar=c=1$ and the metric tensor with the signature $(+---)$. Three- and four-vectors are defined by their components, however, for three-vectors we often use the bold font, for example, $p^\mu = (E, p^1,p^2,p^3) = (E, \pv)$, where  $E = \sqrt{m^2+p^2}$ denotes the particle energy. For the length of a three-vector we use the regular font, $p = |\pv|$. Scalar products of three-vectors are denoted by a dot, $a^\mu b_\mu = a^0 b^0 - \av \cdot \bv$. The unit three-vectors are denoted by a hat so that $\pv = p \, \pvhat$.

%%%%%%%%%%%%%%%%%%%%%%%%%%%%%%%%%%%%%%%%%%%%
\section{center-of-mass (COM) frame} 
\label{sec:COM}

In the analyzes of spin polarization of relativistic particles, it is important to define precisely the reference frames where the specific physical quantities are defined and measured. In this work, we define altogether three different reference frames that are linked by Lorentz boosts and rotations: the center-of-mass frame of the total system, COM, and two rest frames of $\Lambda$'s with a given momentum in COM. The last two frames differ by rotation. 

We assume that the main reference frame corresponds to the center-of-mass frame of the colliding system. In the case of non-central heavy-ion collisions, the axes of the COM frame are defined by the beam axis ($\zvhat$), the impact vector ($\xvhat$), and the direction that is perpendicular to the reaction plane ($\yvhat$) spanned by $\xvhat$ and $\zvhat$, see Fig.~\ref{fig:COMhi}.~\footnote{Here we tacitly assume that the reaction plane angle in the laboratory (LAB) frame can be well determined by calculating the event plane flow vector~\cite{Poskanzer:1998yz}, hence, the COM frame is rotated by this angle around the beam axis in LAB.   The problem that the reaction plane angle is in fact not directly measured is discussed in detail in Sec.~\ref{sec:RP}. } We note that the orientation of the three-vector describing the orbital angular momentum $\Lv$ is opposite to the $y$ axis. 

\begin{figure}[t]
\begin{center}
\includegraphics[width=0.6\textwidth]{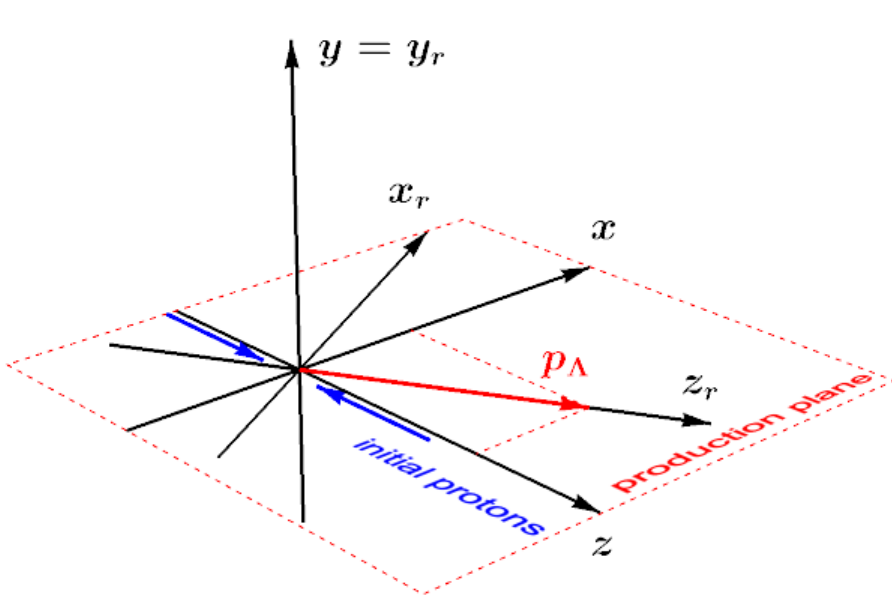}
\end{center}
\caption{The center-of-mass (COM) frame for $p$+$p$ collisions.}
\label{fig:COMpp}
\end{figure}

In the case of proton-proton collisions, the $z$ axis corresponds to the direction of the initial protons, the $y$ axis is defined to be perpendicular to the plane determined by $\zvhat$ and the momentum of the emitted $\Lambda$ hyperon $\pv_\Lambda$, i.e., to the production plane, while $\xvhat$ is perpendicular to both $\yvhat$ and $\zvhat$, see Fig.~\ref{fig:COMpp}. The Cartesian coordinate system $x,y,z$ is taken in the two cases to be right-handed. We note that in the case of proton-proton collisions one may also use a rotated frame where the $z$ axis coincides with the direction of $\pv_\Lambda$. In Fig.~\ref{fig:COMpp} the axes of this frame are denoted by $x_r$, $y_r$, and $z_r$.
\bigskip

%%%%%%%%%%%%%%%%%%%%%%%%%%%%%%%%%%%%%
\section{The \texorpdfstring{$\Lambda$}{Lambda} rest frame \texorpdfstring{$S^\prime(\pvl)$}{S prime}}
\label{sec:Sprime}

In the following, we define two frames where the $\Lambda$ hyperon with the momentum $\pvl$ in COM frame is at rest. The first one is defined by the canonical boost from the COM frame. The second one differs from the first by an additional rotation that aligns the polarization vector with the $z$-axis. 

%%%%%%%%%%%%%%%%%%%%%%%%%%%%%%%%%%%%%
\subsection{The canonical boost}

We define the rest frame $S^\prime(\pvl)$ of $\Lambda$'s with the COM frame three-momentum $\pvl = (p^{1}_{\Lambda}, p^{2}_{\Lambda}, p^{3}_{\Lambda})$ by the canonical boost~\cite{Jackson:1998nia,Leader:2001gr}
\beq
{\cal L}^\mu_{\,\,\,\nu}\left(-{\vv}_\Lambda \right)=\left[\begin{array}{cccc}
\frac{E_{\Lambda}}{m_{\Lambda}} & -\frac{p^{1}_{\small \Lambda}}{m_{\Lambda}} & -\frac{p^{2}_{\Lambda}}{m_{\Lambda}} & -\frac{p^{3}_{\Lambda}}{m_{\Lambda}} \\
-\frac{p^{1}_{\Lambda}}{m_{\Lambda}} & 1+\alpha p^{1}_{\Lambda} p^{1}_{\Lambda} & \alpha p^{1}_{\Lambda} p^{2}_{\Lambda} & \alpha p^{1}_{\Lambda} p^{3}_{\Lambda} \\
-\frac{p^{2}_{\Lambda}}{m_{\Lambda}} & \alpha p^{2}_{\Lambda} p^{1}_{\Lambda} & 1+\alpha p^{2}_{\Lambda} p^{2}_{\Lambda} & \alpha p^{2}_{\Lambda} p^{3}_{\Lambda} \\
-\frac{p^{3}_{\Lambda}}{m_{\Lambda}} & \alpha p^{3}_{\Lambda} p^{1}_{\Lambda} & \alpha p^{3}_{\Lambda} p^{2}_{\Lambda} & 1+\alpha p^{3}_{\Lambda} p^{3}_{\Lambda}
\end{array}\right].
\label{eq:canboost}
\eeq
Here $E_\Lambda$ and $\vv_\Lambda = \pv_\Lambda/E_\Lambda$ are the energy and three-velocity of $\Lambda$ in COM, respectively, while $m_\Lambda$ is the $\Lambda$ mass and $\alpha \equiv 1/\left(m_\Lambda\left( E_\Lambda+m_\Lambda \right)\right)$. We stress that the frame $S^\prime(\pvl)$ depends on $\pvl$ -- in practice one should select an ensemble of events that include $\Lambda$'s with the COM three-momentum in a small bin placed around a given value of $\pvl$. The components of the four-vectors in $S^\prime(\pvl)$ and COM are related by the transformation
\begin{equation}
p^{\prime \mu} =  {\cal L}_{\,\,\,\nu}^{\mu}\left(-{\vv}_\Lambda \right) p^\nu.
\end{equation}
In particular, by construction we obtain $p^{\prime \mu}_\Lambda = (m_\Lambda, 0,0,0)$. 

It is well known~\cite{Leader:2001gr} that the canonical boost can be represented as a superposition of three transformations: the rotation ${\cal R}_\Lambda$ that brings the three-vector $\pv_\Lambda$ to the form $(0,0,p_\Lambda)$, the boost ${\cal L}_{3}$ along the third axis with the velocity $-v_\Lambda$, and the inverse rotation~${\cal R}_\Lambda^{-1}$, namely
\begin{equation}
{\cal L} =  {\cal R}_\Lambda^{-1}(\phi_\Lambda,\theta_\Lambda) {\cal L}_{3}(-v_\Lambda)  {\cal R}_\Lambda(\phi_\Lambda,\theta_\Lambda).
\end{equation}
The rotation ${\cal R}_\Lambda$ can be written as the product of two rotations. If we use the parametrization  $\pv_\Lambda = p_\Lambda\,(
\sin\theta_\Lambda \,\cos\phi_\Lambda,
\sin\theta_\Lambda \,\sin\phi_\Lambda,
\cos\theta_\Lambda)$ in COM, then ${\cal R}_\Lambda =  {\cal R}_{2}(\theta_\Lambda) {\cal R}_{3}(\phi_\Lambda) $, where
\beq
{\cal R}_{3}\left(\phi_\Lambda \right)=\left[\begin{array}{cccc}
1 & 0 & 0 & 0 \\
0 & \cos\phi_\Lambda & \sin\phi_\Lambda &  0 \\
0 & -\sin\phi_\Lambda & \cos\phi_\Lambda &  0 \\
0 & 0 & 0 &  1
\label{eq:rotation}
\end{array}\right]
\eeq
and
\beq
{\cal R}_{2}\left(\theta_\Lambda \right)=\left[\begin{array}{cccc}
1 & 0 & 0 & 0 \\
0 & \cos\theta_\Lambda & 0 & -\sin\theta_\Lambda \\
0 & 0 & 1 & 0 \\
0 & \sin\theta_\Lambda & 0 & \cos\theta_\Lambda
\label{eq:rotation}
\end{array}\right].
\eeq
The boost ${\cal L}_{3}(-v_\Lambda)$ is defined by the expression
\beq
{\cal L}_{3}(-v_\Lambda)
=\left[\begin{array}{cccc}
\gamma_\Lambda & 0  & 0 & -\gamma_\Lambda v_\Lambda \\
0 & 1 & 0 & 0 \\
0 & 0 & 1 & 0\\
-\gamma_\Lambda v_\Lambda  & 0 & 0 & \gamma_\Lambda
\end{array}\right],
\label{eq:canboostz}
\eeq
where $\gamma_\Lambda = E_\Lambda/m_\Lambda$ is the Lorentz factor. Further useful properties of the canonical boost are discussed in Appendix~\ref{sec:propboost}.

%%%%%%%%%%%%%%%%%%%%%%%%%%%%%%%%%%%%%%%%
\subsection{Transformation of the system's angular momentum}

The crucial role in the discussion and interpretation of the spin-polarization measurements is played by the total angular momentum of the system described by the tensor $J^{\mu\nu}$. It can be decomposed into the orbital and spin parts, $J^{\mu\nu}= L^{\mu\nu}+S^{\mu\nu}$. In non-central heavy-ion collisions, a substantial non-zero orbital part $L^{\mu\nu}$ is generated at the initial stage \cite{Becattini:2007sr}. One expects that during the system's evolution some part of $L^{\mu\nu}$ is transferred to the spin part $S^{\mu\nu}$, of course, with the total angular momentum  $J^{\mu\nu}$ being conserved. The generation of a non-zero spin part $S^{\mu\nu}$ may be reflected just by the measured spin polarization of the produced particles. We note that 
the spin part may be also generated at the very early stages of the collision but one expects anyway that the values of $S^{\mu\nu}$ are negligible compared to~$L^{\mu\nu}$.

If one works in the COM frame, only the spatial components of $L^{\mu\nu}$ are different from zero.~\footnote{The conserved quantities $J^{0i}$ corresponding to Lorentz boosts are of the form $J^{0i} = E R^i - t P^i$, where $R^i = (1/E) \int d^3x \,x^i\, T^{00}$ and $E= \int d^3x \,T^{00}$, with $T^{00}$ being the energy density. In the center-of-momentum frame $P^i=0$. Moreover, if the center-of-momentum frame is also the center-of-mass frame (strictly speaking, the center-of-energy for relativistic systems) then we also have $R^i=0$.} They determine the orbital angular momentum of the system through the relation
\begin{equation}
L^k = -\frac{1}{2} \epsilon^{kij} L^{ij}.    
\end{equation}
With the standard orientation of the axes in COM, one expects that the direction of the vector $\Lv$ is opposite to the $y$ axis, see Fig.~\ref{fig:COMhi}. The components of $\Lv$ transform like the components of the magnetic field since they represent spatial components of an antisymmetric tensor $L^{\mu\nu}$. Hence, in the frame $S^\prime(\pvl)$ they are given by the formula~\cite{Jackson:1998nia}
\begin{equation}
\Lv^\prime = \gamma_\Lambda  \Lv
-\frac{\gamma^2_\Lambda}{\gamma_\Lambda + 1}
\, \vv_\Lambda (\vv_\Lambda \cdot \Lv).
\label{eq:Lprime}
\end{equation}
%
% where $\gamma_\Lambda = E_\Lambda/m_\Lambda$. % defined before
From Eq.~(\ref{eq:Lprime}) we find the ratio of the lenghts of the vectors $\Lv^\prime$ and $\Lv$, namely
\begin{equation}
\frac{L^\prime}{L}  = \gamma_\Lambda \left(1 -
(\vv_\Lambda \cdot \Lvhat)^2 \right)^{1/2}. \label{eq:ratio}
\end{equation}

\bigskip

\begin{figure}[t]
\begin{center}
\includegraphics[width=0.4\textwidth]{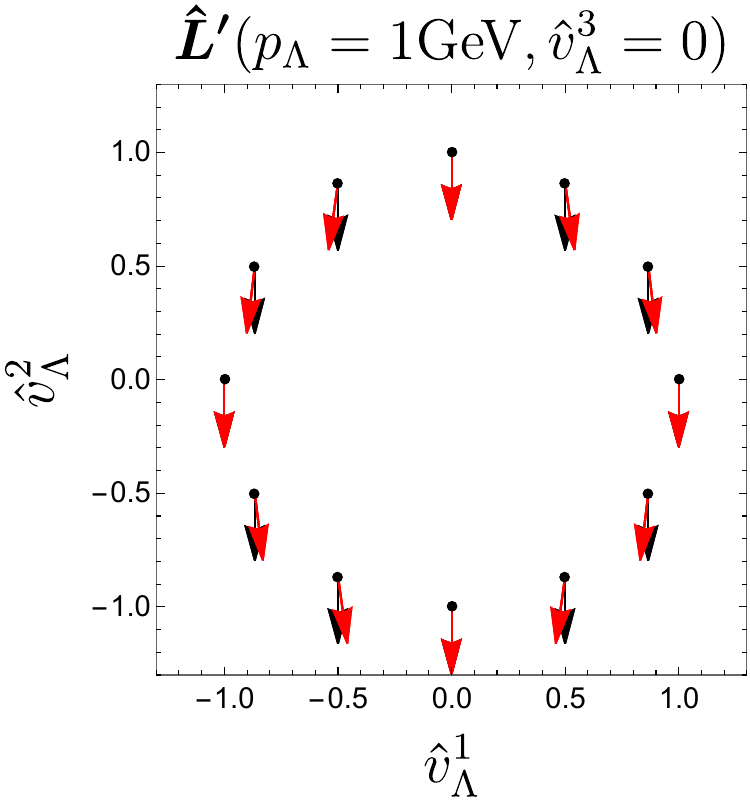}
\includegraphics[width=0.4\textwidth]{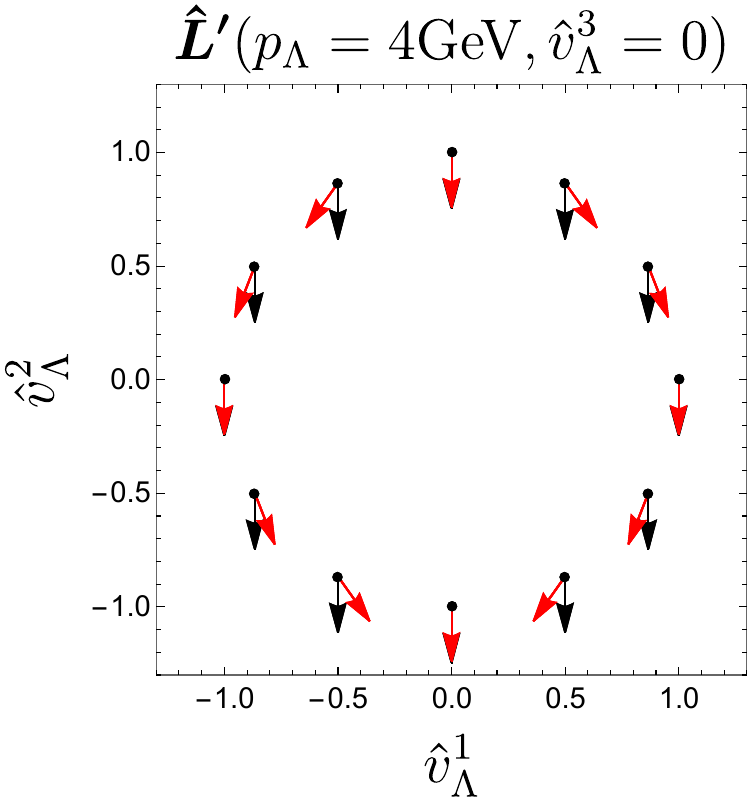}
\end{center}
\caption{ Direction of the total angular momentum of the system transformed from COM to the $S^\prime(\pvl)$ frame (red arrows) for $\Lambda$ particles at midrapidity ($\hat{v}^z=0$) with the momentum $p_\Lambda=1~\text{GeV}$ (left panel) and $p_\Lambda=4~\text{GeV}$ (right panel) for various orientations of the velocity in transverse plane. Black arrows denote the same quantity as seen in COM.} 
\label{fig:Lhat}
\end{figure}

\begin{figure}[t]
\begin{center}
\includegraphics[width=0.4\textwidth]{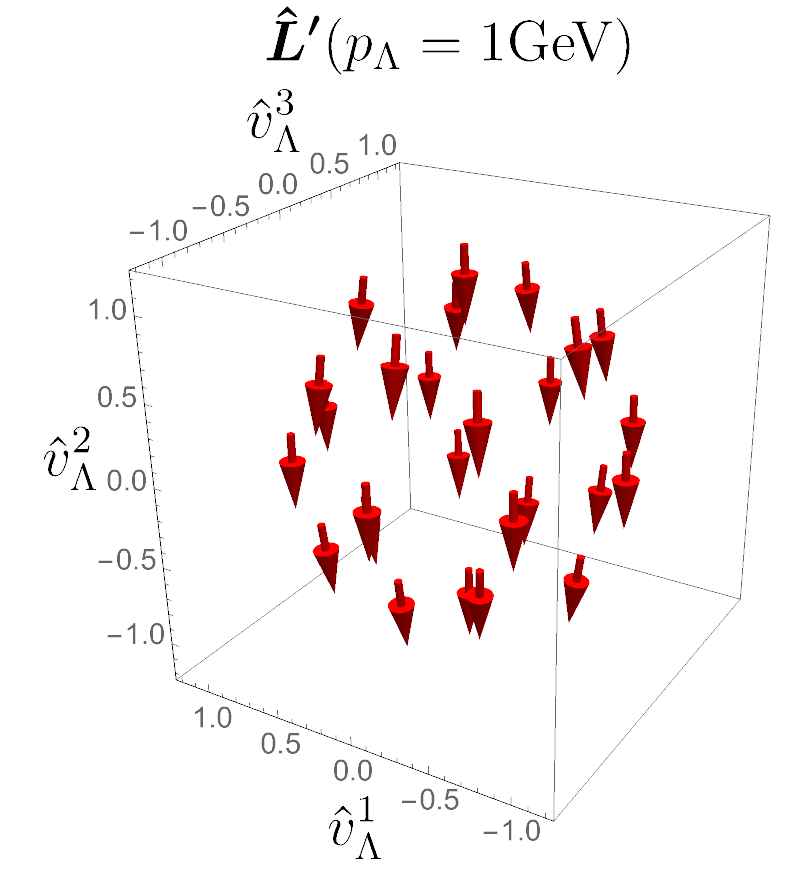}
\includegraphics[width=0.4\textwidth]{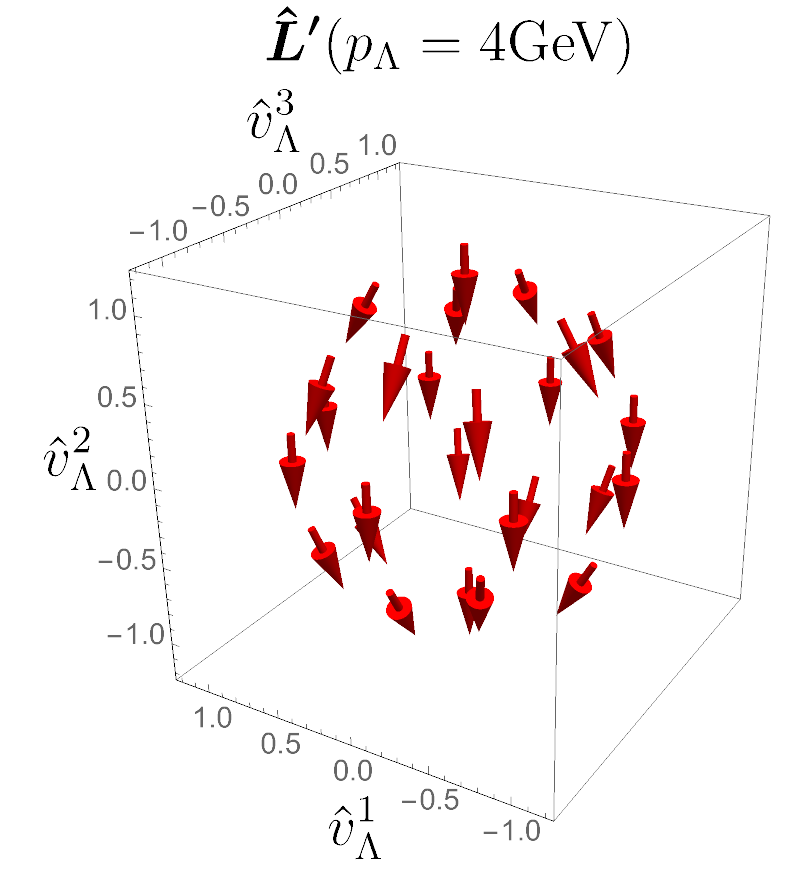}
\end{center}
\caption{Three-dimensional visualization of the vectors $\Lv^\prime$ defined by Eqs.~(\ref{eq:Lprime}) and~(\ref{eq:minus1}).}
\label{fig:Lhat}
\end{figure}

Let us note that for relativistic $\Lambda$'s the directions of $\Lv$ and $\Lv^\prime$ (measured in their appropriate reference frames) may be significantly different. In general, only for the case $\vv_\Lambda \cdot \Lv = 0$ they are the same. For non-relativistic systems, the second term on the right-hand side of Eq.~(\ref{eq:Lprime}) represents a relativistic correction of the order $(\vv_\Lambda/c)^2$ and can be neglected, however, for relativistic systems the second term may be equally important as the first one. Consequently, comparisons of the measured polarization direction should refer to the direction of $\Lv^\prime$ rather than to the direction of $\Lv$. For this purpose, we introduce two unit vectors
\begin{equation}
\Lvhat = \frac{\Lv}{L}, \quad
\Lvhat^\prime = \frac{\Lv^\prime}{L^\prime}.
\end{equation}
Taking into account Eq.~(\ref{eq:ratio}) we may write
\begin{equation}
\Lvhat^\prime =    \left(1 -
(\vv_\Lambda \cdot \Lvhat)^2 \right)^{-1/2} \left(\Lvhat
-\frac{\gamma_\Lambda}{\gamma_\Lambda + 1}
\, \vv_\Lambda (\vv_\Lambda \cdot \Lvhat)
\right).
\label{Lhatprime}
\end{equation}
This vector is expressed only by the three-momentum of $\Lambda$ and the direction of the angular momentum in COM. We note that with our choice of COM,
\begin{equation}
\Lvhat = (0,-1,0)
\label{eq:minus1}
\end{equation}
we obtain
\begin{eqnarray}
\hat{L}^{\prime\, 1} &=& \left(1 -
(v_\Lambda^2)^2 \right)^{-1/2} \frac{\gamma_\Lambda}{\gamma_\Lambda + 1}\,
v_\Lambda^1 v_\Lambda^2, \nonumber \\
\hat{L}^{\prime\, 2} &=& \left(1 -
(v_\Lambda^2)^2 \right)^{-1/2} \left( \frac{\gamma_\Lambda}{\gamma_\Lambda + 1}\,
v_\Lambda^2 v_\Lambda^2 -1 \right), \nonumber \\
\hat{L}^{\prime\, 3} &=& \left(1 -
(v_\Lambda^2)^2 \right)^{-1/2} \frac{\gamma_\Lambda}{\gamma_\Lambda + 1}\,
v_\Lambda^3 v_\Lambda^2.
 \end{eqnarray}
The visualization of those components for the case $v_\Lambda^3 = 0$ is shown in Fig.~\ref{fig:Lhat}.

For the sake of completeness, let us consider the transformation law for the component $K^i = -L^{0i}$ that behaves like an electric-like component of $L^{\mu\nu}$. As we have mentioned above, $\Kv = 0$ in COM. However, after making the canonical boost and using Eq.~(\ref{eq:minus1})  we obtain
\begin{equation}
\Kv^\prime = \gamma_\Lambda \left(v_\Lambda^3,0, -v_\Lambda^1 \right) L  
\end{equation}
or, after normalization,
\begin{equation}
\Kvhat^\prime = \frac{ \left(v_\Lambda^3,0, -v_\Lambda^1 \right)}{\sqrt{(v_\Lambda^1)^2+(v_\Lambda^3)^2}}.  
\end{equation}

\bigskip
\begin{figure}[t]
\begin{center}
\includegraphics[width=0.6\textwidth]{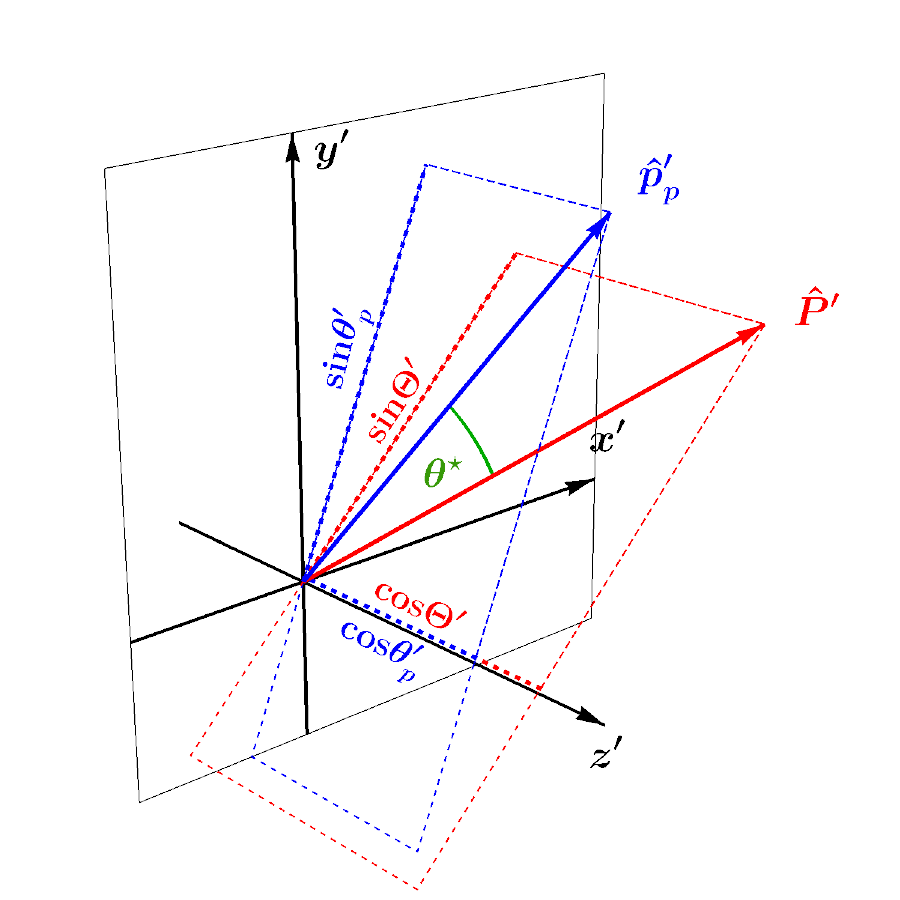}
\end{center}
\caption{The $\Lambda$ rest frame. The momentum distribution of protons produced in the weak decay $\Lambda \to p + \pi^-$ depends on $\cos\theta^*$, where $\theta^*$ is the angle between the polarization vector $\Pv^\prime$ and the proton momentum direction $\pvhat^\prime_p$. }
\label{fig:prime}
\end{figure}

%%%%%%%%%%%%%%%%%%%%%%%%%%%%%
\section{%The $\Lambda$ rest frame $S^*(\pvl)$
The \texorpdfstring{$\Lambda$}{Lambda} rest frame \texorpdfstring{$S^*(\pvl)$}{S star}}
\label{sec:Sstar}

In the $\Lambda$ rest frame $S^\prime(\pvl)$, the $\Lambda$ polarization is characterized by the polarization three-vector $\Pv^\prime$, see Fig.~\ref{fig:prime}. It can be defined by the magnitude $P^\prime$ and the unit vector $\hat{\Pv}^\prime$ that specifies the polarization direction, namely, $\Pv^\prime = P^\prime \hat{\Pv}^\prime$. The vector $\hat{\Pv}^\prime$ can be expressed by the two angles $\Phi^\prime$ and $\Theta^\prime$ with the help of the standard parametrization
\begin{equation}
\hat{\Pv}^\prime = \left(
\sin\Theta^\prime \cos\Phi^\prime,
\sin\Theta^\prime \sin\Phi^\prime,
\cos\Theta^\prime \right).
\end{equation}
In the following, it will be useful to consider also the frame where only the third component of $\hat{\Pv}^\prime$ is different from zero. This is achieved by the subsequent action of the two rotations

\bigskip
\begin{figure}[t]
\begin{center}
\includegraphics[width=0.6\textwidth]{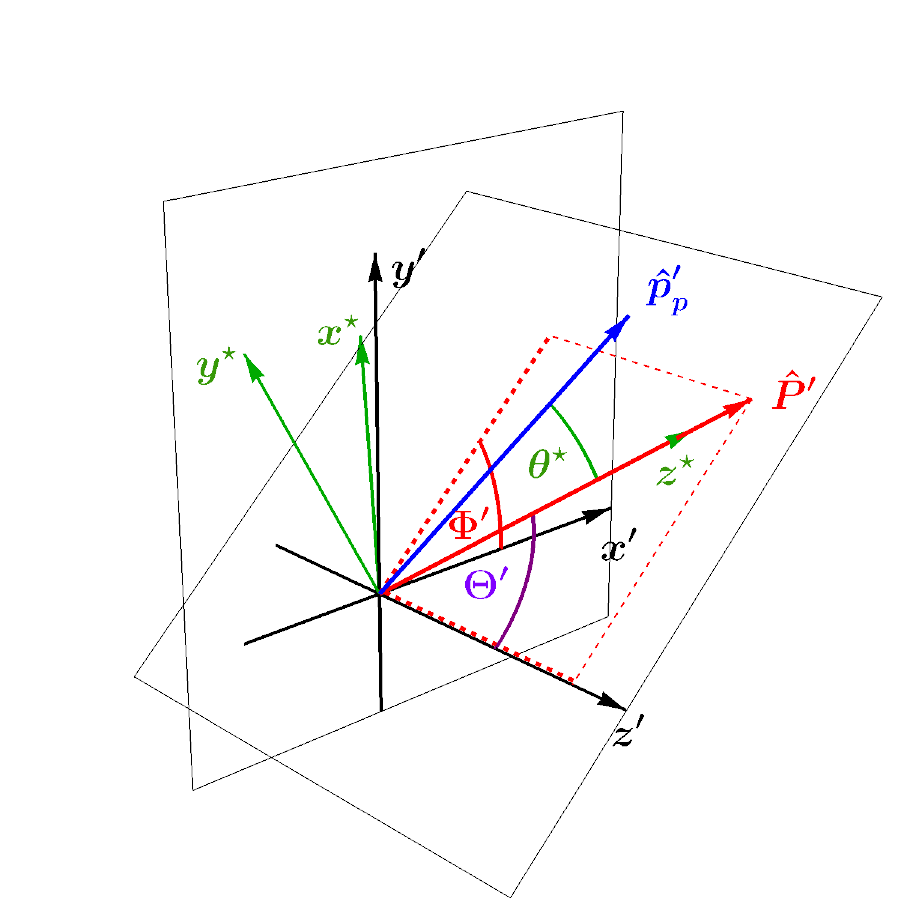}
\end{center}
\caption{The frame $S^*(\pvl)$ is obtained from the frame $S^\prime(\pvl)$ by a rotation that brings $\Pv^\prime$ along the new $z$-axis.}
\label{fig:star}
\end{figure}

\beq
{\cal R}_{z^\prime}\left(\Phi^\prime \right)=\left[\begin{array}{ccc}
\cos\Phi^\prime & \sin\Phi^\prime &  0 \\
-\sin\Phi^\prime & \cos\Phi^\prime &  0 \\
0 & 0 &  1
\label{eq:rotation}
\end{array}\right]
\eeq
and
\beq
{\cal R}_{y^\prime}\left(\Theta^\prime \right)=\left[\begin{array}{ccc}
\cos\Theta^\prime & 0 & -\sin\Theta^\prime \\
0 & 1 & 0 \\
\sin\Theta^\prime & 0 & \cos\Theta^\prime
\label{eq:rotation}
\end{array}\right].
\eeq
The resulting frame will be called $S^*(\pvl)$. It is trivial to see that
\begin{equation}
\hat{\Pv}^* = {\cal R}_{y^\prime}\left(\Theta^\prime \right) {\cal R}_{z^\prime}\left(\Phi^\prime \right) \hat{\Pv}^\prime 
= (0,0,1).
\end{equation}

Let us now consider the three-momentum of the proton emitted in the weak decay of $\Lambda$. Similarly to the case of the polarization vector, we express it as follows $\pv^\prime_p = p_p^\prime\, \hat{\pv}^\prime_p$ where
\begin{equation}
\hat{\pv}^\prime_p =
\left(
\sin\theta^\prime_p \cos\phi^\prime_p,
\sin\theta^\prime_p \sin\phi^\prime_p,
\cos\theta^\prime_p \right).
\end{equation}
In the frame $S^*(\pvl)$ we have
\begin{eqnarray}
\hat{p}^*_{p, x} &=& 
\cos(\Phi^\prime-\phi^\prime_p) \sin\theta^\prime_p \cos\Theta^\prime
-\cos\theta^\prime_p \sin\Theta^\prime\equiv \sin\theta^* \cos\phi^*,
\nonumber \\
\hat{p}^*_{p, y} &=& -\sin(\Phi^\prime-\phi^\prime_p) \sin\theta^\prime_p
\equiv \sin\theta^* \sin\phi^*,
\nonumber \\
\hat{p}^*_{p, z} &=& 
\cos(\Phi^\prime-\phi^\prime_p) \sin\theta^\prime_p\sin\Theta^\prime +\cos\theta^\prime_p \cos\Theta^\prime\equiv \cos\theta^*.
\end{eqnarray}
From the last line, we find~\footnote{This is of course a trivial result. The main reason for introducing the frame $S^*(\pvl)$ is that we find it useful  in the following to consider the angular distributions expressed by the angles $(\theta^\prime_p, \phi^\prime_p)$ or, equivalently, by the angles $(\theta^*, \phi^*)$. }
\begin{equation}
\hat{\Pv}^\prime \cdot \hat{\pv}^\prime_p 
= \hat{\Pv}^* \cdot \hat{\pv}^*_p = \cos\theta^*.
\end{equation}

The angular distributions of protons emitted in the frames $S^\prime(\pvl)$ and $S^*(\pvl)$ satisfy an obvious constraint
\begin{equation}
\int \frac{d N_p}{d \Omega^\prime} \, \sin\theta^\prime_p
d\theta^\prime_p d\phi^\prime_p =
\int \frac{d N_p}{d \Omega^*} \,
\sin\theta^*_p
d\theta^*_p d\phi^*_p,
\end{equation}
where the functions $d N_p/d \Omega$ behave like scalar functions of the azimuthal and polar angles. Consequently, if the distribution $d N_p/d \Omega^*$ is a function of $\cos\theta^*$ only, for example, 
\begin{equation}
\frac{d N_p}{d \Omega^*} = F(\cos\theta^*),   
\end{equation}
where $F(x)$ is an arbitrary function of $x$, then
\begin{equation}
\frac{d N_p}{d \Omega^\prime} = F\left(
\cos(\Phi^\prime-\phi^\prime_p) \sin\theta^\prime_p\sin\Theta^\prime  +\cos\theta^\prime_p\cos\Theta^\prime \right).   
\end{equation}

At the end of this section let us note that the three-vector $\Pv^\prime $ can be interpreted as a spatial part of the four-vector~\cite{Itzykson:1980rh}
\begin{equation}
P^{\prime \, \mu} = \left(0,  \Pv^\prime \right).   
\end{equation}
In general, we have
\begin{equation}
-1 \leq P^\prime \cdot P^\prime \leq 0.
\end{equation}
The case $P^\prime \cdot P^\prime = -1$ corresponds to a pure state in which the spin projection is $1/2$ in the direction of $\hat{\Pv}^\prime$. The values larger than $-1$ describe a mixed spin state, and eventually the value $P^\prime \cdot P^\prime = 0$ means that the system is unpolarized.

%%%%%%%%%%%%%%%%%%%%%%%%%%%%%
\section{The weak decay law}
\label{sec:weakdecay}

In the frame $S^*(\pvl)$, the $\Lambda$ weak decay $\Lambda \to p + \pi^-$ is described by the following law that describes the angular distribution of emitted protons
\beq
\frac{d N^{\rm pol}_p}{d \Omega^{*}}=\frac{1}{4 \pi}\left(1+\alpha_\Lambda \Pv^* \cdot \pvhat_{p}^{*}\right).
\label{eq:weakdecay}
\eeq
Here $\alpha_\Lambda = 0.732$ is the $\Lambda$ decay constant. Equation (\ref{eq:weakdecay}) implies that in $S^\prime(\pvl)$ the proton angular distribution has the form
\beq
\frac{d N^{\rm pol}_p}{d \Omega^\prime} =\frac{1}{4 \pi}\left[1+\alpha_\Lambda P^\prime
\left(
\cos(\Phi^\prime-\phi^\prime_p) \sin\theta^\prime_p\sin\Theta^\prime +\cos\theta^\prime_p\cos\Theta^\prime \right)
\right].
\label{eq:weakdecayp}
\eeq
The averaged values of the three momentum components in $S^\prime(\pvl)$ can be obtained by straightforward integration:
\begin{eqnarray}
\langle \hat{p}^\prime_{p, x} \rangle &=& \int \left( \frac{d N^{\rm pol}_p}{d \Omega^\prime} \right) (\sin\theta^\prime_p)^2 \cos\phi^\prime_p \,  d\theta^\prime_p \, d\phi^\prime_p =
\frac{1}{3} P^\prime \alpha_\Lambda \sin\Theta^\prime 
\cos\Phi^\prime,
\nonumber \\
\langle \hat{p}^\prime_{p, y} \rangle &=& \int \left( \frac{d N^{\rm pol}_p}{d \Omega^\prime} \right) (\sin\theta^\prime_p)^2 \sin\phi^\prime_p \,  d\theta^\prime_p \, d\phi^\prime_p  =
\frac{1}{3} P^\prime \alpha_\Lambda \sin\Theta^\prime 
\sin\Phi^\prime,
\nonumber \\
\langle \hat{p}^\prime_{p, z} \rangle &=& \int \left( \frac{d N^{\rm pol}_p}{d \Omega^\prime} \right) \sin\theta^\prime_p  \cos\theta^\prime_p\,  d\theta^\prime_p \, d\phi^\prime_p =
\frac{1}{3} P^\prime \alpha_\Lambda \cos\Theta^\prime.
\end{eqnarray}
 Here we have introduced angular brackets to denote angular averaging of the proton variables in the Lambda rest frame $S^\prime(\pvl)$. The last result indicates that the magnitude and direction of the polarization can be directly obtained from the averaged values of the three momentum components measured in $S^\prime(\pvl)$
\begin{eqnarray}
\Pv^\prime &=& P^\prime \left(\sin\Theta^\prime 
\cos\Phi^\prime, \sin\Theta^\prime 
\sin\Phi^\prime, \cos\Theta^\prime \right) = \frac{3}{\alpha_\Lambda} \left(\langle \hat{p}^\prime_{p, x} \rangle,
\langle \hat{p}^\prime_{p, y} \rangle,
\langle \hat{p}^\prime_{p, z} \rangle \right).
\label{eq:exp}
\end{eqnarray}
One can also find
\begin{eqnarray}
\langle \cos\phi^\prime_p \rangle &=& \int \left( \frac{d N^{\rm pol}_p}{d \Omega^\prime} \right) \, \sin\theta^\prime_p \cos\phi^\prime_p \,  d\theta^\prime_p \, d\phi^\prime_p =
\frac{\pi \alpha_\Lambda}{8} P^\prime  \sin\Theta^\prime 
\cos\Phi^\prime,
\label{eq:cos} \\
\langle \sin\phi^\prime_p \rangle &=& \int \left( \frac{d N^{\rm pol}_p}{d \Omega^\prime} \right) \sin\theta^\prime_p \sin\phi^\prime_p \,  d\theta^\prime_p \, d\phi^\prime_p  =
\frac{\pi \alpha_\Lambda}{8} P^\prime  \sin\Theta^\prime 
\sin\Phi^\prime.
\label{eq:sine}
\end{eqnarray}
The last expression rewritten in the form 
\begin{equation}
P_H = \frac{8}{\pi \alpha_\Lambda} \langle \sin\phi^\prime_p \rangle  
\end{equation}
serves as the main experimental tool used to determine the spin polarization.~\footnote{The STAR experiment uses the COM frame with the $x$-axis not aligned with the impact vector. In this case, instead of $\langle \sin\phi^\prime_p \rangle$ one studies the mean value of $\langle \sin(\phi^\prime_p -\Psi_{\rm RP}) \rangle$, where $\Psi_{\rm RP}$ denotes the azimuthal angle of the reaction plane, see our discussion in Sec.~\ref{sec:RP}.} The two comments are in order now: 
\begin{itemize}
    \item The quantity $P_H$ is the $y$-component of the polarization three-vector measured in the Lambda rest frame, namely, $P_H = P^\prime  \sin\Theta^\prime \sin\Phi^\prime$. Strictly speaking, it is not the component of the polarization along the total angular momentum vector as the $y$-directions in COM and the Lambda rest frame are different (although the differences for slowly moving Lambdas might be quite small).
    \item In addition to the measurement of the mean $\langle \sin\phi^\prime_p \rangle$ it is tempting to measure, using the same experimental techniques, the mean $\langle \cos\phi^\prime_p \rangle$. Such a measurement would complete the analysis of the three components of the polarization vector in the Lambda rest frame, as the longitudinal component has been already measured. The ratio of such measurements would give us directly the information about the angle $\Phi^\prime$.
\end{itemize}

At this point, it is convenient to discuss the effect of the detector magnetic field on the spin polarization. If the $\Lambda$ hyperons move in a magnetic field, their spins undergo precession with the frequency $\omega^\prime = g_\Lambda \, (\mu_N/\hbar) \, B^\prime$. Here $g_\Lambda = 0.613$ is the magnitude of the Land\'e $g$-factor for $\Lambda$'s, $\mu_N$ is the nuclear magneton, and $B^\prime$ is the magnitude of the magnetic field in the Lambda rest frame. If $\Lambda$'s move perpendicularly to the magnetic field, the field in the rest frame equals $B^\prime = \gamma_\Lambda B = \gamma_\Lambda\, x\, T$, where we have expressed the value of the magnetic field in COM in units of Tesla, $B = x\, T$. The mean angle by which the spin direction changes equals $\Delta \Phi_{\rm prec} = \omega^\prime \Delta t^\prime$, where $\Delta t^\prime = 2.63 \cdot 10^{-10}$~s is the mean Lambda lifetime in its rest frame. This altogether gives $\Delta \Phi_{\rm prec} = 0.0077 \,x\, \gamma_\Lambda$. In our opinion, this value represents the systematic error for the experimental estimates of the angles $\Theta^\prime$ and $\Phi^\prime$. We note that the present STAR estimate of $\Delta \Phi_{\rm prec}$ is somewhat larger, namely, gives 0.022 for $x=0.5$ and $\gamma_\Lambda \sim 2$. Nevertheless, this value is used to argue that the effect of the spin precession on the global polarization measurements
is negligible. From the point of our analysis, the effects of spin precession require more detailed studies where the impact of precession on the estimates of the angles $\Theta^\prime$ and $\Phi^\prime$ can be clarified.

%%%%%%%%%%%%%%%%%%%%%%%%%%%%%%%%%%%%%%%%%%%% 
\section{Correlation with total angular momentum} 
\label{sec:master}

\subsection{Improved formula for the projection}

We have discussed above how the magnitude and direction of the spin polarization can be determined in the frame where $\Lambda$'s are at rest. More precisely, we have considered the rest frame of $\Lambda$'s with three-momentum $\pv_\Lambda$, which is obtained by the canonical boost from COM. A natural question at this stage appears, how the direction of the measured polarization is related to the axes of the COM coordinate system. 

Equation (\ref{eq:exp}) gives the prescription how to measure three independent components of the $\Lambda$ polarization in its (canonical) rest frame. Assuming that the measurement of the averages $\langle \hat{p}^\prime_{p, x} \rangle$, $\langle \hat{p}^\prime_{p, y} \rangle$, and $\langle \hat{p}^\prime_{p, z} \rangle $ is indeed possible, we may define the projection of the polarization along the direction of the total angular momentum by the expression
\begin{equation}
\Lvhat^\prime \cdot \Pv^\prime =    \left(1 -
(\vv_\Lambda \cdot \Lvhat)^2 \right)^{-1/2} 
\left(\Lvhat \cdot \Pv^\prime
-\frac{\gamma_\Lambda}{\gamma_\Lambda + 1}
\, \vv_\Lambda \cdot \Pv^\prime \,\,\, \vv_\Lambda \cdot \Lvhat
\right).
\label{eq:master}
\end{equation}
The direction represented by a unit vector $\Lvhat^\prime$ is the direction of the total angular momentum that is ``seen'' by the spin of the decaying $\Lambda$ that has three-momentum $\pv_\Lambda$ in COM. By construction $|\Lvhat^\prime \cdot \Pv^\prime| \leq P^\prime \leq 1$.

The measurement of the $\Lambda$ spin polarization is very often interpreted as an analog of the Einstein-de Haas and/or the Barnett effect \cite{dehaas:1915,Barnett:1935}. Except for the fact that these two phenomena describe the behavior of a different physical system, one important difference is that for these two effects there exists always a reference frame  where all particles are at rest~\footnote{Although it is typically a non-inertial rotating frame, the non-relativistic treatment allows for simple addition of polarizations of different particles.}. In the case of spin polarization of $\Lambda$'s, such a frame does not exist, since the analyzed $\Lambda$'s have usually different momenta in COM. 

So far, our discussion has been concentrated on $\Lambda$'s with a given momentum in COM. For a given colliding system, beam energy, and the centrality class, such $\Lambda$'s can be treated as produced in the same physical environment (even if they are ``taken'' from different events) so it makes sense to obtain $\Lvhat \cdot \Pv^\prime$ or $\Lvhat^\prime \cdot \Pv^\prime$ from the proton distributions in $S^\prime(\pv_\Lambda)$. The advantage of the expression (\ref{eq:master}) compared to the estimate of just $\Lvhat \cdot \Pv^\prime$ is that the spin polarization of each $\Lambda$, irrespectively of its three-momentum $\pv_\Lambda$ in COM, is projected on the same physical axis corresponding to $\Lv$ in COM. Hence, Eq.~(\ref{eq:master}) is in our opinion the proper object that can be used to study the relation between the polarization of all $\Lambda$'s with $\Lv$. To do so, one has to simply average 
Eq.~(\ref{eq:master}) over all $\Lambda$'s with different $\pv_\Lambda$.

\subsection{Numerical estimate of the relativistic effects}

To make a numerical estimate of the effects discussed above, we consider the case where $\Pv^\prime = P^\prime \Lvhat$ and
\begin{equation}
\Lvhat^\prime \cdot \Pv^\prime =  P^\prime   \left(1 -
v_2^2 \right)^{-1/2} 
\left(1
-\frac{v_2^2}{1 + \sqrt{1-v^2}}
\, 
\right) \equiv P^\prime F_P(\vv)
\label{eq:masterN}
\end{equation}
where $(v_1,v_2,v_3)$ are the components of the $\Lambda$ velocity in COM and $v= \sqrt{v_1^2+v_2^2+v_3^2}$ (to simplify the notation we skip here the subscript $\Lambda$). We further assume that the velocity distribution of Lambdas is thermal and described by the Fermi-Dirac distribution
\begin{equation} 
F_T(v) = N \left[\exp\left( \frac{m_\lambda}{T_{\rm eff} \sqrt{1-v^2}} \right)+1 \right]^{-1}.
\end{equation}
Here $T_{\rm eff}$ is an effective temperature and $N$ is the normalization constant that is irrelevant for our study. The average value of $\Lvhat^\prime \cdot \Pv^\prime$ for Lambdas with the momentum in the range between $m$~GeV and $n$~GeV is defined as the ratio
\begin{equation}
\langle \Lvhat^\prime \cdot \Pv^\prime \rangle_{m-n}     = P^\prime \,\,
\frac{\int_{v_{(m)}}^{v_{(n)}} dv \int d\Omega  \,F_P(\vv) F_T(v)}{\int_{v_{(m)}}^{v_{(n)}} dv \int d\Omega \, F_T(v)},
\end{equation}
where
\begin{equation}
v_{(n)} = \tanh\left[\sinh^{-1}\left(\frac{n \, \rm GeV}{m_\Lambda}\right)\right].   
\end{equation}
The numerical calculations performed with $T_{\rm eff}$~=~150~MeV give: 
\mbox{$\langle\Lvhat^\prime \cdot \Pv^\prime \rangle_{2-3}= 0.97~P^\prime$},\mbox{ 
$\langle\Lvhat^\prime\cdot\Pv^\prime\rangle_{3-4}= 0.94~P^\prime$}, $\langle \Lvhat^\prime \cdot \Pv^\prime \rangle_{4-5}= 0.92~P^\prime$, and $\langle \Lvhat^\prime \cdot \Pv^\prime \rangle_{5-6}= 0.90~P^\prime$. Consequently, the relativistic effects studied in this work may reach 10\% for the most energetic Lambdas studied at STAR. However, in the case of momentum-averaged results in the range $0.5-6 \mathrm{GeV} / c$ studied in Ref.~\cite{STAR:2017ckg} one obtains a negligible correction $\langle\hat{\boldsymbol{L}}^{\prime} \cdot \boldsymbol{P}^{\prime}\rangle_{0.5-6}=0.997 P^{\prime}$.

%%%%%%%%%%%%%%%%%%%%%%%%%%%%%%%%%%%%%%%%%%%%%%%%
\subsection{Replacing the reaction plane angle by the experimentally-determined event plane angle}
\label{sec:RP} 

Equation (\ref{eq:master}) is an algebraic equation involving $\Lvhat$, which makes the experimental determination of $\Lvhat^\prime \cdot \Pv^\prime$ more difficult than the measurement of $\Lvhat \cdot \Pv^\prime$ alone. This is due to the fact that in the general case the reaction plane is characterized by the angle $\PsiRP$ that is not necessarily equal to zero and the direction of the orbital angular momentum is defined by the vector
\begin{equation}
 \Lvhat = -\left( \cos\left(\PsiRP+\pi/2\right),\sin\left(\PsiRP+\pi/2\right),0\right).
 \label{LPsiRP}
\end{equation}
We note that for $\PsiRP=0$ this formula is reduced to Eq.~(\ref{eq:minus1}) that has been used so far.

To measure $\Lvhat \cdot \Pv^\prime$ in the case $\PsiRP \neq 0$ we first perform averaging over the angles of the emitted protons and get
\begin{eqnarray}
 \Lvhat \cdot \Pv^\prime = -\frac{8}{\pi \alpha_\Lambda} \langle \sin(\phi^\prime_p-\PsiRP) \rangle = -\frac{8}{\pi \alpha_\Lambda} \int 
\left( \frac{d N^{\rm pol}_p}{d \Omega^\prime} \right)  \sin\left( \phi^\prime_p-\PsiRP  \right) \,  d \Omega^{\prime}.
\end{eqnarray}
This equation introduces an explicit dependence of our results on the reaction plane angle $\PsiRP$ which is not directly measured. To overcome this difficulty, the STAR experiment measures the azimuthal angle of the event plane, $\PsiEP$, determined by the directed flow. Consequently, one considers the observable
\begin{eqnarray}
&& \langle \sin(\phi^\prime_p-\PsiEP) \rangle_{\rm ev.} =
\langle \sin(\phi^\prime_p-\PsiRP-(\PsiEP-\PsiRP)) \rangle_{\rm ev.} \nonumber \\
&& = \langle \sin(\phi^\prime_p-\PsiRP) \rangle_{\rm ev.}
\langle \cos \DPsi \rangle_{\rm ev.}
\equiv  \langle \sin(\phi^\prime_p-\PsiRP) \rangle_{\rm ev.}  R^{(1)}_{\rm EP} ,
\label{resolution1}
\end{eqnarray}
where we introduced $\DPsi \equiv\PsiEP-\PsiRP$. In Eq.~(\ref{resolution1}) one assumes that $\phi^\prime_p$ and $\PsiEP$ are correlated only with the reaction plane angle $\PsiRP$ and uses the property $\langle \sin \DPsi  \rangle_{\rm ev.} = 0$. The notation $\langle ... \rangle_{\rm ev.}$ means that one makes first averaging over different protons in one event and then makes averaging over a sample of events. The last equation in (\ref{resolution1}) defines the reaction plane resolution factor $R^{(1)}_{\rm EP}$. Using Eq.~(\ref{resolution1}) we find the formula 
\begin{equation}
\langle \Lvhat \cdot \Pv^\prime \rangle_{\rm ev.} = -\frac{8}{\pi \alpha_\Lambda} \langle \sin(\phi^\prime_p-\PsiRP) \rangle_{\rm ev.} 
= -\frac{8}{\pi \alpha_\Lambda}  \frac{
\langle \sin(\phi^\prime_p-\PsiEP) \rangle_{\rm ev.}}
{R^{(1)}_{\rm EP}},
\end{equation}
which is the basis of the experimental approach (for example, see Eq. (19) in Ref.~\cite{Becattini:2020ngo}).

The method used for the measurement of $\Lvhat \cdot \Pv^\prime$ suggests a treatment of Eq.~(\ref{eq:master}). Since the expected modifications are at the level of 10\% we can make an expansion of the right-hand side of Eq.~(\ref{eq:master}) in powers of $v_\Lambda/c$. Up to quadratic terms we obtain
\begin{equation}
\Lvhat^\prime \cdot \Pv^\prime = 
\Lvhat \cdot \Pv^\prime
-\frac{\gamma_\Lambda}{\gamma_\Lambda + 1}
\, \vv_\Lambda \cdot \Pv^\prime \,\,\, \vv_\Lambda \cdot \Lvhat
+ \frac{1}{2} (\vv_\Lambda \cdot \Lvhat)^2 \Lvhat \cdot \Pv^\prime.
\label{eq:masterapprox}
\end{equation}
Consequently, our task is reduced to the determination of the two additional averages: $\langle \vv_\Lambda \cdot \Pv^\prime \,\,\, \vv_\Lambda \cdot \Lvhat \rangle_{\rm ev.}$ and $\langle (\vv_\Lambda \cdot \Lvhat)^2 \Lvhat \cdot \Pv^\prime \rangle_{\rm ev.}$.

For the $\Lambda$ hyperons produced at midrapidity (i.e., for $\theta_\Lambda=\pi/2$) we may use the property
\begin{eqnarray}
 \langle \vv_\Lambda \cdot \Pv^\prime
\,\,\, \vv_\Lambda \cdot \Lvhat \rangle_{\rm ev.} = \frac{8}{\pi \alpha_\Lambda} \langle \cos(\phi_\Lambda- \phi^\prime_p) \sin(\phi_\Lambda-\PsiRP)  \rangle_{\rm ev.}.
\end{eqnarray}
To replace the dependence on $\PsiRP$ by the dependence on $\PsiEP$, we use the equation
\begin{equation}
\langle \cos(\phi_\Lambda- \phi^\prime_p) \sin(\phi_\Lambda-\PsiRP)  \rangle_{\rm ev.} 
= \frac{\langle \cos(\phi_\Lambda- \phi^\prime_p) \sin(\phi_\Lambda-\PsiEP)  \rangle_{\rm ev.} }{\langle \cos(\phi_\Lambda- \phi^\prime_p) \cos \DPsi  \rangle_{\rm ev.} }.
\end{equation}
where we used the assumptions that $\langle \cos(\phi_\Lambda- \phi^\prime_p) \sin\DPsi  \rangle_{\rm ev.} = 0$. The denominator in the last equation can be treated as another resolution parameter. 

%%%%%%%%%%%%
In the case of the term $\langle (\vv_\Lambda \cdot \Lvhat)^2 \Lvhat \cdot \Pv^\prime \rangle_{\rm ev.}$ we use the following property
\begin{equation}
\langle (\vv_\Lambda \cdot \Lvhat)^2 \Lvhat \cdot \Pv^\prime \rangle_{\rm ev.} =
-\frac{8}{\pi \alpha_\Lambda} 
\langle \sin^2(\phi_\Lambda-\PsiRP)  \sin(\phi^\prime_p-\PsiRP) 
\rangle_{\rm ev.}.
\end{equation}
Following the same steps as above (i.e., assuming that the averages of the odd functions of $\DPsi$ vanish) we may construct two observables
\begin{eqnarray}
\langle \sin^2(\phi_\Lambda-\PsiEP) \sin(\phi^\prime_p-\PsiEP) \rangle_{\rm ev.}  &=& M_1
\langle \cos\DPsi \cos (2\DPsi) \rangle_{\rm ev.}  \nonumber \\
&& \hspace{-5cm} + \,( M_2 \,+\, M_3)
\langle \cos\DPsi \sin^2 \DPsi \rangle_{\rm ev.}
\label{eqs_for_M_1}
\end{eqnarray}
and
\begin{eqnarray}
  \langle \sin (2(\phi_\Lambda-\PsiEP))
\cos(\phi^\prime_p-\PsiEP) \rangle_{\rm ev.}  &=& \left( M_2 +2(M_3 - 2M_1) \right)
\langle \cos \DPsi \cos (2\DPsi) \rangle_{\rm ev.}  
\nonumber \\
&& \hspace{-5cm} - \,2(M_3-2 M_1) 
\langle \cos ^3 \DPsi \rangle_{\rm ev.} ,
\label{eqs_for_M_2}
\end{eqnarray}
where
\begin{eqnarray}
M_1 &=& \langle \sin^2(\phi_\Lambda-\PsiRP)
\sin(\phi^\prime_p-\PsiRP) \rangle_{\rm ev.}, \nonumber \\
M_2 &=& \langle \sin \left( 2(\phi_\Lambda-\PsiRP) \right)
\cos(\phi^\prime_p-\PsiRP)  \rangle_{\rm ev.}, 
\nonumber \\
M_3 &=& \langle \sin(\phi^\prime_p-\PsiRP) \rangle_{\rm ev.} . \label{eq:Ms}
\end{eqnarray}
Since the quantity $M_3$ can be measured (see our analysis of Eq.~(\ref{resolution1})) Eqs.~(\ref{eqs_for_M_1}) and (\ref{eqs_for_M_2}) allow for the determination of the quantities $M_1$ and $M_2$, provided the left-hand sides of Eqs.~(\ref{eqs_for_M_1}) and (\ref{eqs_for_M_2}), as well as $\langle \cos ^3\!\DPsi \rangle_{\rm ev.}$, are measurable \footnote{Note that $\langle \cos\DPsi \cos (2(\DPsi)) \rangle_{\rm ev.}=-R^{(1)}_{\rm EP}+2\langle \cos ^3 \DPsi \rangle_{\rm ev.}$ and $\langle \cos\DPsi \sin^2 \DPsi \rangle_{\rm ev.}=R^{(1)}_{\rm EP}-\langle \cos ^3 \DPsi \rangle_{\rm ev.}$.}. Since the quantity $\langle (\vv_\Lambda \cdot \Lvhat)^2 \Lvhat \cdot \Pv^\prime \rangle_{\rm ev.}$ is directly expressed by $M_1$, it can be also measured.

%%%%%%%%%%%%%%%%%%%%%%%%%%%%%%%%%%%%%%
\section{Proton-proton collisions}

At the end of this work, let us turn to a discussion of polarization measurement in proton-proton collisions \cite{Bunce:1976yb,Panagiotou:1989sv}. If the proton-proton COM frame corresponds to the case shown in Fig. 2, where the variant with a rotation in the production plane is chosen, the canonical boost is reduced to the form (\ref{eq:canboostz}). Then, the four-vector describing the polarization in COM is obtained by the boost ${\cal L}_3\left(+v_\Lambda \right)$ acting on the four-vector $(0, \Pv^\prime)$. This leads to the expression
\begin{equation}
P^\mu = P^\prime \left(
\gamma_\Lambda v_\Lambda \cos\Theta^\prime, \sin\Theta^\prime \cos\Phi^\prime, \sin\Theta^\prime \sin\Phi^\prime,
\gamma_\Lambda \cos\Theta^\prime
\right).
\label{eq:pp}
\end{equation}
At first sight, the interpretation of the spin polarization measurements in proton-proton collisions seems to be easier compared to the heavy-ion case. As the transverse components of $\Pv^\prime$ are not affected by the boost one may try simply to add them and average over different $\Lambda$'s. This procedure, however, makes sense only if the $x$ and $z$ components of $\Pv^\prime$ are zero. Otherwise, the results obtained for different $\Lambda$'s depend on the boost and the original transition to a rotated frame. 

Consequently, if the spin polarization of $\Lambda$'s has non-zero  $x$ and $z$ components it is suitable to use the frame without the rotation. In this case, we may follow the procedure discussed above for heavy ions, with the total angular momentum direction replaced by one of the other physical directions defined in the non-rotated COM frame that can be measured (for example, the direction perpendicular to the plane determined by the beam and the fastest proton produced). Such a procedure may be also useful in the case if more $\Lambda$'s are produced in one event.

%%%%%%%%%%%%%%%%%%%%%%%%%%%%%%%%%%%%%%%%%%%%
\section{Conclusions} 
\label{sec:conclusions}

In this work, we have discussed the interpretation of the recent measurements of the spin polarization of $\Lambda$ hyperons produced in relativistic heavy-ion collisions. We have shown that the precise interpretation of the relation between the $\Lambda$ spin direction (measured in the $\Lambda$ rest frame) and the total angular momentum of the system (measured in the center-of-mass frame) requires that the direction of the angular momentum is boosted to the $\Lambda$ rest frame. We have given the necessary formula that, we hope, may find its practical implementation in the polarization measurements. In particular, this expression may be used to average the measured polarization of $\Lambda$'s with different momenta in the center-of-mass frame. Several explicit expressions for boosts and rotations have been written out, which may help to compare model predictions with the experimental results. 

\bigskip
%%%%%%%%%%%%%%%%%%%%%%%%%%%%%%%%%%%%%%%%%%%%
{\it Acknowledgements.} We thank Y. Bondar, M. Ga\'zdzicki, T. Niida, I. Selyuzhenkov, G. Stefanek, and S. Voloshin for stimulating and clarifying discussions. The work of WF and RR was supported in part by the Polish National Science Center Grants No.~2016/23/B/ST2/00717 and No.~2018/30/E/ST2/00432, respectively.
%%%%%%%%%%%%%%%%%%%%%%%%%%%%%%%%%%%%%%%%%%%% 
 
\begin{appendix}

%%%%%%%%%%%%%%%%%%%%%%%%%%%%%
\section{Properties of the canonical boost}
\label{sec:propboost}

Of course, the three-momentum of the $\Lambda$ hyperon in its frame is zero, however, it is possible to introduce the four-vector in COM that defines the direction of a moving $\Lambda$ and has nonvanishing components in the $\Lambda$ rest frame. The desired object is
\begin{equation}
\lambda_\parallel^\mu =  (\lambda_\parallel^0,\lamv_\parallel)  = \left(0, \hat{\pv}_\Lambda \right). 
\label{eq:lampar}
\end{equation}
After the canonical boost to $S^\prime(\pvl)$ we obtain
\begin{equation}
\lambda^{\prime \,\mu}_\parallel =
(\lambda_\parallel^{\prime \, 0},\lamv^\prime_\parallel) =
\gamma_\Lambda\left( -\frac{p_\Lambda}{E_\Lambda}, \hat{\pv}_\Lambda\right). 
\end{equation}
The property $\lamv^\prime_\parallel = \gamma_\Lambda \lamv_\parallel$ is usually interpreted as the conservation of the angles between the three-momentum of a moving particle and the frame axes by the canonical boost, as one has $\lamv^\prime_\parallel / \lambda^\prime_\parallel =\lamv_\parallel / \lambda_\parallel$. The other two important four-vectors are:
\begin{equation}
\lambda_{\perp, 1}^\mu =  (\lambda_{\perp, 1}^0,\lamv_{\perp,1})  = \frac{1}{\sqrt{(p_{\Lambda}^1)^2+(p_{\Lambda}^2)^2}} \left(0, -p_{\Lambda}^2, p_{\Lambda}^1, 0 \right) 
\label{eq:lamper1}
\end{equation}
and
\begin{equation}
\lambda_{\perp, 2}^\mu =  (\lambda_{\perp, 2}^0,\lamv_{\perp, 2}) = \frac{1}{p_\Lambda \sqrt{(p_{\Lambda}^1)^2+(p_{\Lambda}^2)^2}} \left(0, -p_{\Lambda}^3 p_{\Lambda}^1, -p_{\Lambda}^3 p_{\Lambda}^2, (p_{\Lambda}^1)^2+(p_{\Lambda}^2)^2 \right). 
\label{eq:lamper2}
\end{equation}
The four-vectors $\lambda_{\perp, 1}^\mu$ and $\lambda_{\perp, 2}^\mu$ do not change under the canonical boost~(\ref{eq:canboost}). The three-vector $\lamv_{\perp,1}$ represents the rotation axis for the rotation ${\cal R}_\Lambda$.

For example, any four-vector of the form
\begin{equation}
n^\mu = (0, \nv) = (0, n^1, n^2, n^3),  
\end{equation}
with the normalization $n^\mu n_\mu = -1$ or, equivalently, $\nv \cdot \nv =1$, after the canonical boost we obtain %
\begin{equation}
n^{\prime \, \mu} = \left(n^{\prime \,0}, \nv^\prime \right) =
\left(
-\frac{\pv_\Lambda \cdot \nv}{m_\Lambda}, \,
\nv + \pv_\Lambda \,\, \frac{\pv_\Lambda \cdot \nv}{m_\Lambda 
\left( E_\Lambda+m_\Lambda \right)} \right),
\end{equation}
where $\nv^\prime \cdot \nv^\prime = 1 + (\pv_\Lambda \cdot \nv)^2/m_\Lambda^2$. Thus the direction $\nv$ in COM can be 

One can check that
\begin{equation}
\lamv^\prime_\parallel \cdot \nv^\prime = \gamma_\Lambda^2 \,
\lamv_\parallel \cdot \nv \, .
\end{equation}   
Since $n^0=0$ and $P^{\prime \, 0}=0$ we obtain
\begin{equation}
\nv \cdot \Pv = \nv^\prime \cdot \Pv^\prime. 
\end{equation}
Hence, the four-vector $n^\mu$ can be used to define the polarization direction in the way that is frame independent. 

%%%%%%%%%%%%%%%%%%%%%%%%%%%%%
\section{Distribution of the proton three-momenta along an arbitrary direction.}

If the distribution of protons coming from the $\Lambda$ decay is given by Eq.~(\ref{eq:weakdecay}), their angular distribution in $S^\prime(\pvl)$ is obtained from Eq.~(\ref{eq:weakdecayp}). In this section, we assume that a certain angular distribution of protons  $dN_p/d\Omega^\prime$ is known and construct the distribution of the proton projected momentum along an arbitrary direction in $S^\prime(\pvl)$. The obtained formula can be used to determine polarization in a given direction directly from the angular distribution $dN_p/d\Omega^\prime$. Note that if the distribution $dN_p/d\Omega^\prime$ is isotropic, the proton projected momentum along any direction has a flat distribution that reflects no sign of polarization.

We start with the integral of the angular distribution and rewrite as follows (we are now in the frame $S^\prime(\pvl)$ but for clarity of notation we skip the index prime, also the number of protons is normalized to one)
\begin{equation}
1 = \int \left(\frac{dN}{d\Omega}\right) \sin\theta_p d\theta_p d\phi_p = \int\limits_{-1}^{+1} dc \int \left(\frac{dN}{d\Omega}\right) 
\delta(c^* - c) \sin\theta_p d\theta_p d\phi_p.
\label{eq:1}
\end{equation}
Here $\delta$ denotes the Dirac delta function and $c^*$ is the cosine of the angle between the proton direction defined by the angles $\theta_p$ and $\phi_p$ and an arbitrary direction defined by the angles $\Theta$ and $\Phi$, hence
\begin{equation}
c^* = \cos\theta^* = \cos\Theta \cos\theta_p + \cos(\Phi-\phi_p)  \sin\Theta \sin\theta_p.   
\end{equation}
By construction $-1 \leq c^* \leq +1$. 

The distribution of the proton three-momentum direction along the direction specified by the angles $\Theta$ and $\Phi$ is defined by the integral
\begin{equation}
\frac{dN}{dc}  =  \int \left(\frac{dN}{d\Omega}\right) 
\delta(c^* - c) \sin\theta_p d\theta_p d\phi_p.
\label{eq:dNc1}
\end{equation}
To do the integral on the right-hand side we introduce the function
\begin{equation}
f(c,\Theta,\Phi,\theta_p,\phi_p) = c^*(\Theta,\Phi,\theta_p,\phi_p)-c  
\end{equation}
and use the properties of the Dirac delta function to write
\begin{equation}
\frac{dN}{dc}  =  \int\limits_0^\pi 
\sin\theta_p d\theta_p
\int\limits_0^{2\pi}
\frac{dN}{d\Omega}(\theta_p,\phi_p)
\left[ 
\frac{\delta(\phi_p - \phi_p^+)}{|f'(\phi_p^+)|}
+ \frac{\delta(\phi_p - \phi_p^-)}{|f'(\phi_p^-)|)}
\right]
 d\phi_p.
\label{eq:dNc2}
\end{equation}
Here
\begin{equation}
f^\prime =  \sin(\Phi-\phi_p)  \sin\Theta \sin\theta_p  
\end{equation}
and
\begin{equation}
\phi_p^\pm = \Phi \pm \arccos \left(\frac{c -  \cos\Theta \cos\theta_p}{\sin\Theta 
\sin\theta_p } \right)  
\label{eq:phipm}
\end{equation}
are the two solutions of the equation $c^*-c = 0$. We note that it may happen that the solutions defined by Eq.~(\ref{eq:phipm}) are outside of the range $(0,2\pi)$, however, since $f^\prime$ and $dN/d\Omega$ are periodic this does not lead to problems. As a matter of fact, this equation has solutions only if the following condition is satisfied 
\begin{equation}
-1 \leq \frac{c -  \cos\Theta \cos\theta_p}{\sin\Theta 
\sin\theta_p } \leq +1.
\label{eq:cond}
\end{equation}
This implies that the range of the integration over $\theta_p$ must be limited --- for given values of $c$ and $\Theta$, only those values of $\theta_p$ contribute to the integral (\ref{eq:dNc2}) which satisfy (\ref{eq:cond}). If we introduce the notation $c =\cos\theta$, with $0 \leq \theta \leq \pi$, then the limits for the $\theta_p$ integration are
\begin{equation}
\theta_p^{\rm min} = \max(0,\Theta-\theta,\theta-\Theta)
\,\leq \theta_p\,
\,\leq \,
\min(\pi,2 \pi - \theta -\Theta,\Theta+\theta)
=\theta_p^{\rm max}.
\end{equation}
Here we assumed that the range of the angles $\Theta$ and $\theta_p$ is between $0$ and $\pi$. Consequently, the final result can be written as
\begin{equation}
\frac{dN}{dc}  =  \int\limits_{\theta_p^{\rm min}}^{\theta_p^{\rm max}} 
\left[ \frac{dN}{d\Omega}(\theta_p,\phi_p^+)
\frac{1}{|f'(\phi_p^+)|}
+ 
\frac{dN}{d\Omega}(\theta_p,\phi_p^-)
\frac{1}{|f'(\phi_p^-)|)}
\right] \, \sin\theta_p d\theta_p.
\label{eq:dNc3}
\end{equation}
If the proton distribution is given by the weak decay law discussed above, the last formula may be interpreted as the inverse of the transformation that leads from Eq.~(\ref{eq:weakdecay}) to Eq.~(\ref{eq:weakdecayp}). We have checked numerically that this is indeed so.

\end{appendix}

\newpage
%%%%%%%%%%%%%%%%%%%%%%%%%%%%%%%%%%%%%%%%%
%\bibliographystyle{utphys} 

\bibliography{spinref}{}

\begin{thebibliography}{61}
\expandafter\ifx\csname natexlab\endcsname\relax\def\natexlab#1{#1}\fi
\expandafter\ifx\csname bibnamefont\endcsname\relax
  \def\bibnamefont#1{#1}\fi
\expandafter\ifx\csname bibfnamefont\endcsname\relax
  \def\bibfnamefont#1{#1}\fi
\expandafter\ifx\csname citenamefont\endcsname\relax
  \def\citenamefont#1{#1}\fi
\expandafter\ifx\csname url\endcsname\relax
  \def\url#1{\texttt{#1}}\fi
\expandafter\ifx\csname urlprefix\endcsname\relax\def\urlprefix{URL }\fi
\providecommand{\bibinfo}[2]{#2}
\providecommand{\eprint}[2][]{\url{#2}}

\bibitem[{\citenamefont{Bunce et~al.}(1976)}]{Bunce:1976yb}
\bibinfo{author}{\bibfnamefont{G.}~\bibnamefont{Bunce}} \bibnamefont{et~al.},
  \bibinfo{journal}{Phys. Rev. Lett.} \textbf{\bibinfo{volume}{36}},
  \bibinfo{pages}{1113} (\bibinfo{year}{1976}).

\bibitem[{\citenamefont{Bourrely et~al.}(1980)\citenamefont{Bourrely, Leader,
  and Soffer}}]{BOURRELY198095}
\bibinfo{author}{\bibfnamefont{C.}~\bibnamefont{Bourrely}},
  \bibinfo{author}{\bibfnamefont{E.}~\bibnamefont{Leader}}, \bibnamefont{and}
  \bibinfo{author}{\bibfnamefont{J.}~\bibnamefont{Soffer}},
  \bibinfo{journal}{Physics Reports} \textbf{\bibinfo{volume}{59}},
  \bibinfo{pages}{95} (\bibinfo{year}{1980}), ISSN \bibinfo{issn}{0370-1573},
  \urlprefix\url{https://www.sciencedirect.com/science/article/pii/0370157380900174}.

\bibitem[{\citenamefont{Panagiotou}(1990)}]{Panagiotou:1989sv}
\bibinfo{author}{\bibfnamefont{A.~D.} \bibnamefont{Panagiotou}},
  \bibinfo{journal}{Int. J. Mod. Phys. A} \textbf{\bibinfo{volume}{5}},
  \bibinfo{pages}{1197} (\bibinfo{year}{1990}).

\bibitem[{\citenamefont{Becattini et~al.}(2021)\citenamefont{Becattini, Liao,
  and Lisa}}]{Becattinibook}
\bibinfo{author}{\bibfnamefont{F.}~\bibnamefont{Becattini}},
  \bibinfo{author}{\bibfnamefont{J.}~\bibnamefont{Liao}}, \bibnamefont{and}
  \bibinfo{author}{\bibfnamefont{M.}~\bibnamefont{Lisa}},
  \bibinfo{journal}{Lect. Notes Phys.} \textbf{\bibinfo{volume}{987}}
  (\bibinfo{year}{2021}).

\bibitem[{\citenamefont{Jacob and Rafelski}(1987)}]{Jacob:1987sj}
\bibinfo{author}{\bibfnamefont{M.}~\bibnamefont{Jacob}} \bibnamefont{and}
  \bibinfo{author}{\bibfnamefont{J.}~\bibnamefont{Rafelski}},
  \bibinfo{journal}{Phys. Lett.} \textbf{\bibinfo{volume}{B190}},
  \bibinfo{pages}{173} (\bibinfo{year}{1987}).

\bibitem[{\citenamefont{Anikina et~al.}(1984)}]{Anikina:1984cu}
\bibinfo{author}{\bibfnamefont{M.~K.} \bibnamefont{Anikina}}
  \bibnamefont{et~al.}, \bibinfo{journal}{Z. Phys.}
  \textbf{\bibinfo{volume}{C25}}, \bibinfo{pages}{1} (\bibinfo{year}{1984}).

\bibitem[{\citenamefont{Bartke et~al.}(1990)}]{Bartke:1990cn}
\bibinfo{author}{\bibfnamefont{J.}~\bibnamefont{Bartke}} \bibnamefont{et~al.}
  (\bibinfo{collaboration}{NA35}), \bibinfo{journal}{Z. Phys.}
  \textbf{\bibinfo{volume}{C48}}, \bibinfo{pages}{191} (\bibinfo{year}{1990}).

\bibitem[{\citenamefont{Liang and Wang}(2005)}]{Liang:2004ph}
\bibinfo{author}{\bibfnamefont{Z.-T.} \bibnamefont{Liang}} \bibnamefont{and}
  \bibinfo{author}{\bibfnamefont{X.-N.} \bibnamefont{Wang}},
  \bibinfo{journal}{Phys. Rev. Lett.} \textbf{\bibinfo{volume}{94}},
  \bibinfo{pages}{102301} (\bibinfo{year}{2005}), \bibinfo{note}{[Erratum:
  Phys. Rev. Lett.96,039901(2006)]}, \eprint{nucl-th/0410079}.

\bibitem[{\citenamefont{Betz et~al.}(2007)\citenamefont{Betz, Gyulassy, and
  Torrieri}}]{Betz:2007kg}
\bibinfo{author}{\bibfnamefont{B.}~\bibnamefont{Betz}},
  \bibinfo{author}{\bibfnamefont{M.}~\bibnamefont{Gyulassy}}, \bibnamefont{and}
  \bibinfo{author}{\bibfnamefont{G.}~\bibnamefont{Torrieri}},
  \bibinfo{journal}{Phys. Rev.} \textbf{\bibinfo{volume}{C76}},
  \bibinfo{pages}{044901} (\bibinfo{year}{2007}), \eprint{0708.0035}.

\bibitem[{\citenamefont{Voloshin}(2004)}]{Voloshin:2004ha}
\bibinfo{author}{\bibfnamefont{S.~A.} \bibnamefont{Voloshin}}
  (\bibinfo{year}{2004}), \eprint{nucl-th/0410089}.

\bibitem[{\citenamefont{Abelev et~al.}(2007)}]{Abelev:2007zk}
\bibinfo{author}{\bibfnamefont{B.~I.} \bibnamefont{Abelev}}
  \bibnamefont{et~al.} (\bibinfo{collaboration}{STAR}), \bibinfo{journal}{Phys.
  Rev.} \textbf{\bibinfo{volume}{C76}}, \bibinfo{pages}{024915}
  (\bibinfo{year}{2007}), \bibinfo{note}{[Erratum: Phys.
  Rev.C95,no.3,039906(2017)]}, \eprint{0705.1691}.

\bibitem[{\citenamefont{Becattini and Piccinini}(2008)}]{Becattini:2007nd}
\bibinfo{author}{\bibfnamefont{F.}~\bibnamefont{Becattini}} \bibnamefont{and}
  \bibinfo{author}{\bibfnamefont{F.}~\bibnamefont{Piccinini}},
  \bibinfo{journal}{Annals Phys.} \textbf{\bibinfo{volume}{323}},
  \bibinfo{pages}{2452} (\bibinfo{year}{2008}), \eprint{0710.5694}.

\bibitem[{\citenamefont{Becattini et~al.}(2008)\citenamefont{Becattini,
  Piccinini, and Rizzo}}]{Becattini:2007sr}
\bibinfo{author}{\bibfnamefont{F.}~\bibnamefont{Becattini}},
  \bibinfo{author}{\bibfnamefont{F.}~\bibnamefont{Piccinini}},
  \bibnamefont{and} \bibinfo{author}{\bibfnamefont{J.}~\bibnamefont{Rizzo}},
  \bibinfo{journal}{Phys. Rev.} \textbf{\bibinfo{volume}{C77}},
  \bibinfo{pages}{024906} (\bibinfo{year}{2008}), \eprint{0711.1253}.

\bibitem[{\citenamefont{Becattini
  et~al.}(2013{\natexlab{a}})\citenamefont{Becattini, Chandra, Del~Zanna, and
  Grossi}}]{Becattini:2013fla}
\bibinfo{author}{\bibfnamefont{F.}~\bibnamefont{Becattini}},
  \bibinfo{author}{\bibfnamefont{V.}~\bibnamefont{Chandra}},
  \bibinfo{author}{\bibfnamefont{L.}~\bibnamefont{Del~Zanna}},
  \bibnamefont{and} \bibinfo{author}{\bibfnamefont{E.}~\bibnamefont{Grossi}},
  \bibinfo{journal}{Annals Phys.} \textbf{\bibinfo{volume}{338}},
  \bibinfo{pages}{32} (\bibinfo{year}{2013}{\natexlab{a}}), \eprint{1303.3431}.

\bibitem[{\citenamefont{Becattini
  et~al.}(2013{\natexlab{b}})\citenamefont{Becattini, Csernai, and
  Wang}}]{Becattini:2013vja}
\bibinfo{author}{\bibfnamefont{F.}~\bibnamefont{Becattini}},
  \bibinfo{author}{\bibfnamefont{L.}~\bibnamefont{Csernai}}, \bibnamefont{and}
  \bibinfo{author}{\bibfnamefont{D.~J.} \bibnamefont{Wang}},
  \bibinfo{journal}{Phys. Rev.} \textbf{\bibinfo{volume}{C88}},
  \bibinfo{pages}{034905} (\bibinfo{year}{2013}{\natexlab{b}}),
  \bibinfo{note}{[Erratum: Phys. Rev.C93,no.6,069901(2016)]},
  \eprint{1304.4427}.

\bibitem[{\citenamefont{Becattini et~al.}(2017)\citenamefont{Becattini,
  Karpenko, Lisa, Upsal, and Voloshin}}]{Becattini:2016gvu}
\bibinfo{author}{\bibfnamefont{F.}~\bibnamefont{Becattini}},
  \bibinfo{author}{\bibfnamefont{I.}~\bibnamefont{Karpenko}},
  \bibinfo{author}{\bibfnamefont{M.}~\bibnamefont{Lisa}},
  \bibinfo{author}{\bibfnamefont{I.}~\bibnamefont{Upsal}}, \bibnamefont{and}
  \bibinfo{author}{\bibfnamefont{S.}~\bibnamefont{Voloshin}},
  \bibinfo{journal}{Phys. Rev.} \textbf{\bibinfo{volume}{C95}},
  \bibinfo{pages}{054902} (\bibinfo{year}{2017}), \eprint{1610.02506}.

\bibitem[{\citenamefont{Karpenko and Becattini}(2017)}]{Karpenko:2016jyx}
\bibinfo{author}{\bibfnamefont{I.}~\bibnamefont{Karpenko}} \bibnamefont{and}
  \bibinfo{author}{\bibfnamefont{F.}~\bibnamefont{Becattini}},
  \bibinfo{journal}{Eur. Phys. J. C} \textbf{\bibinfo{volume}{77}},
  \bibinfo{pages}{213} (\bibinfo{year}{2017}), \eprint{1610.04717}.

\bibitem[{\citenamefont{Li et~al.}(2017)\citenamefont{Li, Pang, Wang, and
  Xia}}]{Li:2017slc}
\bibinfo{author}{\bibfnamefont{H.}~\bibnamefont{Li}},
  \bibinfo{author}{\bibfnamefont{L.-G.} \bibnamefont{Pang}},
  \bibinfo{author}{\bibfnamefont{Q.}~\bibnamefont{Wang}}, \bibnamefont{and}
  \bibinfo{author}{\bibfnamefont{X.-L.} \bibnamefont{Xia}},
  \bibinfo{journal}{Phys. Rev. C} \textbf{\bibinfo{volume}{96}},
  \bibinfo{pages}{054908} (\bibinfo{year}{2017}), \eprint{1704.01507}.

\bibitem[{\citenamefont{Xie et~al.}(2017)\citenamefont{Xie, Wang, and
  Csernai}}]{Xie:2017upb}
\bibinfo{author}{\bibfnamefont{Y.}~\bibnamefont{Xie}},
  \bibinfo{author}{\bibfnamefont{D.}~\bibnamefont{Wang}}, \bibnamefont{and}
  \bibinfo{author}{\bibfnamefont{L.~P.} \bibnamefont{Csernai}},
  \bibinfo{journal}{Phys. Rev. C} \textbf{\bibinfo{volume}{95}},
  \bibinfo{pages}{031901} (\bibinfo{year}{2017}), \eprint{1703.03770}.

\bibitem[{\citenamefont{Sun and Ko}(2017)}]{Sun:2017xhx}
\bibinfo{author}{\bibfnamefont{Y.}~\bibnamefont{Sun}} \bibnamefont{and}
  \bibinfo{author}{\bibfnamefont{C.~M.} \bibnamefont{Ko}},
  \bibinfo{journal}{Phys. Rev. C} \textbf{\bibinfo{volume}{96}},
  \bibinfo{pages}{024906} (\bibinfo{year}{2017}), \eprint{1706.09467}.

\bibitem[{\citenamefont{Adamczyk et~al.}(2017)}]{STAR:2017ckg}
\bibinfo{author}{\bibfnamefont{L.}~\bibnamefont{Adamczyk}} \bibnamefont{et~al.}
  (\bibinfo{collaboration}{STAR}), \bibinfo{journal}{Nature}
  \textbf{\bibinfo{volume}{548}}, \bibinfo{pages}{62} (\bibinfo{year}{2017}),
  \eprint{1701.06657}.

\bibitem[{\citenamefont{Adam et~al.}(2018)}]{Adam:2018ivw}
\bibinfo{author}{\bibfnamefont{J.}~\bibnamefont{Adam}} \bibnamefont{et~al.}
  (\bibinfo{collaboration}{STAR}), \bibinfo{journal}{Phys. Rev. C}
  \textbf{\bibinfo{volume}{98}}, \bibinfo{pages}{014910}
  (\bibinfo{year}{2018}), \eprint{1805.04400}.

\bibitem[{\citenamefont{Acharya et~al.}(2020)}]{Acharya:2019vpe}
\bibinfo{author}{\bibfnamefont{S.}~\bibnamefont{Acharya}} \bibnamefont{et~al.}
  (\bibinfo{collaboration}{ALICE}), \bibinfo{journal}{Phys. Rev. Lett.}
  \textbf{\bibinfo{volume}{125}}, \bibinfo{pages}{012301}
  (\bibinfo{year}{2020}), \eprint{1910.14408}.

\bibitem[{\citenamefont{Gao et~al.}(2008)\citenamefont{Gao, Chen, Deng, Liang,
  Wang, and Wang}}]{Gao:2007bc}
\bibinfo{author}{\bibfnamefont{J.-H.} \bibnamefont{Gao}},
  \bibinfo{author}{\bibfnamefont{S.-W.} \bibnamefont{Chen}},
  \bibinfo{author}{\bibfnamefont{W.-T.} \bibnamefont{Deng}},
  \bibinfo{author}{\bibfnamefont{Z.-T.} \bibnamefont{Liang}},
  \bibinfo{author}{\bibfnamefont{Q.}~\bibnamefont{Wang}}, \bibnamefont{and}
  \bibinfo{author}{\bibfnamefont{X.-N.} \bibnamefont{Wang}},
  \bibinfo{journal}{Phys. Rev.} \textbf{\bibinfo{volume}{C77}},
  \bibinfo{pages}{044902} (\bibinfo{year}{2008}), \eprint{0710.2943}.

\bibitem[{\citenamefont{Chen et~al.}(2009)\citenamefont{Chen, Deng, Gao, and
  Wang}}]{Chen:2008wh}
\bibinfo{author}{\bibfnamefont{S.-W.} \bibnamefont{Chen}},
  \bibinfo{author}{\bibfnamefont{J.}~\bibnamefont{Deng}},
  \bibinfo{author}{\bibfnamefont{J.-H.} \bibnamefont{Gao}}, \bibnamefont{and}
  \bibinfo{author}{\bibfnamefont{Q.}~\bibnamefont{Wang}},
  \bibinfo{journal}{Front. Phys. China} \textbf{\bibinfo{volume}{4}},
  \bibinfo{pages}{509} (\bibinfo{year}{2009}), \eprint{0801.2296}.

\bibitem[{\citenamefont{Becattini and Tinti}(2010)}]{Becattini:2009wh}
\bibinfo{author}{\bibfnamefont{F.}~\bibnamefont{Becattini}} \bibnamefont{and}
  \bibinfo{author}{\bibfnamefont{L.}~\bibnamefont{Tinti}},
  \bibinfo{journal}{Annals Phys.} \textbf{\bibinfo{volume}{325}},
  \bibinfo{pages}{1566} (\bibinfo{year}{2010}), \eprint{0911.0864}.

\bibitem[{\citenamefont{Becattini}(2012)}]{Becattini:2012tc}
\bibinfo{author}{\bibfnamefont{F.}~\bibnamefont{Becattini}},
  \bibinfo{journal}{Phys. Rev. Lett.} \textbf{\bibinfo{volume}{108}},
  \bibinfo{pages}{244502} (\bibinfo{year}{2012}), \eprint{1201.5278}.

\bibitem[{\citenamefont{Becattini and Grossi}(2015)}]{Becattini:2015nva}
\bibinfo{author}{\bibfnamefont{F.}~\bibnamefont{Becattini}} \bibnamefont{and}
  \bibinfo{author}{\bibfnamefont{E.}~\bibnamefont{Grossi}},
  \bibinfo{journal}{Phys. Rev.} \textbf{\bibinfo{volume}{D92}},
  \bibinfo{pages}{045037} (\bibinfo{year}{2015}), \eprint{1505.07760}.

\bibitem[{\citenamefont{Hayata et~al.}(2015)\citenamefont{Hayata, Hidaka,
  Noumi, and Hongo}}]{Hayata:2015lga}
\bibinfo{author}{\bibfnamefont{T.}~\bibnamefont{Hayata}},
  \bibinfo{author}{\bibfnamefont{Y.}~\bibnamefont{Hidaka}},
  \bibinfo{author}{\bibfnamefont{T.}~\bibnamefont{Noumi}}, \bibnamefont{and}
  \bibinfo{author}{\bibfnamefont{M.}~\bibnamefont{Hongo}},
  \bibinfo{journal}{Phys. Rev.} \textbf{\bibinfo{volume}{D92}},
  \bibinfo{pages}{065008} (\bibinfo{year}{2015}), \eprint{1503.04535}.

\bibitem[{\citenamefont{Florkowski
  et~al.}(2018{\natexlab{a}})\citenamefont{Florkowski, Friman, Jaiswal, and
  Speranza}}]{Florkowski:2017ruc}
\bibinfo{author}{\bibfnamefont{W.}~\bibnamefont{Florkowski}},
  \bibinfo{author}{\bibfnamefont{B.}~\bibnamefont{Friman}},
  \bibinfo{author}{\bibfnamefont{A.}~\bibnamefont{Jaiswal}}, \bibnamefont{and}
  \bibinfo{author}{\bibfnamefont{E.}~\bibnamefont{Speranza}},
  \bibinfo{journal}{Phys. Rev.} \textbf{\bibinfo{volume}{C97}},
  \bibinfo{pages}{041901} (\bibinfo{year}{2018}{\natexlab{a}}),
  \eprint{1705.00587}.

\bibitem[{\citenamefont{Florkowski
  et~al.}(2018{\natexlab{b}})\citenamefont{Florkowski, Friman, Jaiswal,
  Ryblewski, and Speranza}}]{Florkowski:2017dyn}
\bibinfo{author}{\bibfnamefont{W.}~\bibnamefont{Florkowski}},
  \bibinfo{author}{\bibfnamefont{B.}~\bibnamefont{Friman}},
  \bibinfo{author}{\bibfnamefont{A.}~\bibnamefont{Jaiswal}},
  \bibinfo{author}{\bibfnamefont{R.}~\bibnamefont{Ryblewski}},
  \bibnamefont{and} \bibinfo{author}{\bibfnamefont{E.}~\bibnamefont{Speranza}},
  \bibinfo{journal}{Phys. Rev. D} \textbf{\bibinfo{volume}{97}},
  \bibinfo{pages}{116017} (\bibinfo{year}{2018}{\natexlab{b}}),
  \eprint{1712.07676}.

\bibitem[{\citenamefont{Florkowski
  et~al.}(2019{\natexlab{a}})\citenamefont{Florkowski, Kumar, Ryblewski, and
  Mazeliauskas}}]{Florkowski:2019voj}
\bibinfo{author}{\bibfnamefont{W.}~\bibnamefont{Florkowski}},
  \bibinfo{author}{\bibfnamefont{A.}~\bibnamefont{Kumar}},
  \bibinfo{author}{\bibfnamefont{R.}~\bibnamefont{Ryblewski}},
  \bibnamefont{and}
  \bibinfo{author}{\bibfnamefont{A.}~\bibnamefont{Mazeliauskas}},
  \bibinfo{journal}{Phys. Rev. C} \textbf{\bibinfo{volume}{100}},
  \bibinfo{pages}{054907} (\bibinfo{year}{2019}{\natexlab{a}}),
  \eprint{1904.00002}.

\bibitem[{\citenamefont{Li et~al.}(2021)\citenamefont{Li, Stephanov, and
  Yee}}]{Li:2020eon}
\bibinfo{author}{\bibfnamefont{S.}~\bibnamefont{Li}},
  \bibinfo{author}{\bibfnamefont{M.~A.} \bibnamefont{Stephanov}},
  \bibnamefont{and} \bibinfo{author}{\bibfnamefont{H.-U.} \bibnamefont{Yee}},
  \bibinfo{journal}{Phys. Rev. Lett.} \textbf{\bibinfo{volume}{127}},
  \bibinfo{pages}{082302} (\bibinfo{year}{2021}), \eprint{2011.12318}.

\bibitem[{\citenamefont{Hu}(2021)}]{Hu:2021lnx}
\bibinfo{author}{\bibfnamefont{J.}~\bibnamefont{Hu}}, \bibinfo{journal}{Phys.
  Rev. D} \textbf{\bibinfo{volume}{103}}, \bibinfo{pages}{116015}
  (\bibinfo{year}{2021}), \eprint{2101.08440}.

\bibitem[{\citenamefont{Gao et~al.}(2012)\citenamefont{Gao, Liang, Pu, Wang,
  and Wang}}]{Gao:2012ix}
\bibinfo{author}{\bibfnamefont{J.-H.} \bibnamefont{Gao}},
  \bibinfo{author}{\bibfnamefont{Z.-T.} \bibnamefont{Liang}},
  \bibinfo{author}{\bibfnamefont{S.}~\bibnamefont{Pu}},
  \bibinfo{author}{\bibfnamefont{Q.}~\bibnamefont{Wang}}, \bibnamefont{and}
  \bibinfo{author}{\bibfnamefont{X.-N.} \bibnamefont{Wang}},
  \bibinfo{journal}{Phys. Rev. Lett.} \textbf{\bibinfo{volume}{109}},
  \bibinfo{pages}{232301} (\bibinfo{year}{2012}), \eprint{1203.0725}.

\bibitem[{\citenamefont{Chen et~al.}(2013)\citenamefont{Chen, Pu, Wang, and
  Wang}}]{Chen:2012ca}
\bibinfo{author}{\bibfnamefont{J.-W.} \bibnamefont{Chen}},
  \bibinfo{author}{\bibfnamefont{S.}~\bibnamefont{Pu}},
  \bibinfo{author}{\bibfnamefont{Q.}~\bibnamefont{Wang}}, \bibnamefont{and}
  \bibinfo{author}{\bibfnamefont{X.-N.} \bibnamefont{Wang}},
  \bibinfo{journal}{Phys. Rev. Lett.} \textbf{\bibinfo{volume}{110}},
  \bibinfo{pages}{262301} (\bibinfo{year}{2013}), \eprint{1210.8312}.

\bibitem[{\citenamefont{Fang et~al.}(2016)\citenamefont{Fang, Pang, Wang, and
  Wang}}]{Fang:2016vpj}
\bibinfo{author}{\bibfnamefont{R.-H.} \bibnamefont{Fang}},
  \bibinfo{author}{\bibfnamefont{L.-G.} \bibnamefont{Pang}},
  \bibinfo{author}{\bibfnamefont{Q.}~\bibnamefont{Wang}}, \bibnamefont{and}
  \bibinfo{author}{\bibfnamefont{X.-N.} \bibnamefont{Wang}},
  \bibinfo{journal}{Phys. Rev.} \textbf{\bibinfo{volume}{C94}},
  \bibinfo{pages}{024904} (\bibinfo{year}{2016}), \eprint{1604.04036}.

\bibitem[{\citenamefont{Fang et~al.}(2017)\citenamefont{Fang, Pang, Wang, and
  Wang}}]{Fang:2016uds}
\bibinfo{author}{\bibfnamefont{R.-H.} \bibnamefont{Fang}},
  \bibinfo{author}{\bibfnamefont{J.-Y.} \bibnamefont{Pang}},
  \bibinfo{author}{\bibfnamefont{Q.}~\bibnamefont{Wang}}, \bibnamefont{and}
  \bibinfo{author}{\bibfnamefont{X.-N.} \bibnamefont{Wang}},
  \bibinfo{journal}{Phys. Rev.} \textbf{\bibinfo{volume}{D95}},
  \bibinfo{pages}{014032} (\bibinfo{year}{2017}), \eprint{1611.04670}.

\bibitem[{\citenamefont{Florkowski
  et~al.}(2018{\natexlab{c}})\citenamefont{Florkowski, Kumar, and
  Ryblewski}}]{Florkowski:2018ahw}
\bibinfo{author}{\bibfnamefont{W.}~\bibnamefont{Florkowski}},
  \bibinfo{author}{\bibfnamefont{A.}~\bibnamefont{Kumar}}, \bibnamefont{and}
  \bibinfo{author}{\bibfnamefont{R.}~\bibnamefont{Ryblewski}},
  \bibinfo{journal}{Phys. Rev. C} \textbf{\bibinfo{volume}{98}},
  \bibinfo{pages}{044906} (\bibinfo{year}{2018}{\natexlab{c}}),
  \eprint{1806.02616}.

\bibitem[{\citenamefont{Weickgenannt et~al.}(2019)\citenamefont{Weickgenannt,
  Sheng, Speranza, Wang, and Rischke}}]{Weickgenannt:2019dks}
\bibinfo{author}{\bibfnamefont{N.}~\bibnamefont{Weickgenannt}},
  \bibinfo{author}{\bibfnamefont{X.-L.} \bibnamefont{Sheng}},
  \bibinfo{author}{\bibfnamefont{E.}~\bibnamefont{Speranza}},
  \bibinfo{author}{\bibfnamefont{Q.}~\bibnamefont{Wang}}, \bibnamefont{and}
  \bibinfo{author}{\bibfnamefont{D.~H.} \bibnamefont{Rischke}},
  \bibinfo{journal}{Phys. Rev. D} \textbf{\bibinfo{volume}{100}},
  \bibinfo{pages}{056018} (\bibinfo{year}{2019}), \eprint{1902.06513}.

\bibitem[{\citenamefont{Weickgenannt et~al.}(2021)\citenamefont{Weickgenannt,
  Speranza, Sheng, Wang, and Rischke}}]{Weickgenannt:2020aaf}
\bibinfo{author}{\bibfnamefont{N.}~\bibnamefont{Weickgenannt}},
  \bibinfo{author}{\bibfnamefont{E.}~\bibnamefont{Speranza}},
  \bibinfo{author}{\bibfnamefont{X.-l.} \bibnamefont{Sheng}},
  \bibinfo{author}{\bibfnamefont{Q.}~\bibnamefont{Wang}}, \bibnamefont{and}
  \bibinfo{author}{\bibfnamefont{D.~H.} \bibnamefont{Rischke}},
  \bibinfo{journal}{Phys. Rev. Lett.} \textbf{\bibinfo{volume}{127}},
  \bibinfo{pages}{052301} (\bibinfo{year}{2021}), \eprint{2005.01506}.

\bibitem[{\citenamefont{Bhadury
  et~al.}(2021{\natexlab{a}})\citenamefont{Bhadury, Florkowski, Jaiswal, Kumar,
  and Ryblewski}}]{Bhadury:2020puc}
\bibinfo{author}{\bibfnamefont{S.}~\bibnamefont{Bhadury}},
  \bibinfo{author}{\bibfnamefont{W.}~\bibnamefont{Florkowski}},
  \bibinfo{author}{\bibfnamefont{A.}~\bibnamefont{Jaiswal}},
  \bibinfo{author}{\bibfnamefont{A.}~\bibnamefont{Kumar}}, \bibnamefont{and}
  \bibinfo{author}{\bibfnamefont{R.}~\bibnamefont{Ryblewski}},
  \bibinfo{journal}{Phys. Lett. B} \textbf{\bibinfo{volume}{814}},
  \bibinfo{pages}{136096} (\bibinfo{year}{2021}{\natexlab{a}}),
  \eprint{2002.03937}.

\bibitem[{\citenamefont{Bhadury
  et~al.}(2021{\natexlab{b}})\citenamefont{Bhadury, Florkowski, Jaiswal, Kumar,
  and Ryblewski}}]{Bhadury:2020cop}
\bibinfo{author}{\bibfnamefont{S.}~\bibnamefont{Bhadury}},
  \bibinfo{author}{\bibfnamefont{W.}~\bibnamefont{Florkowski}},
  \bibinfo{author}{\bibfnamefont{A.}~\bibnamefont{Jaiswal}},
  \bibinfo{author}{\bibfnamefont{A.}~\bibnamefont{Kumar}}, \bibnamefont{and}
  \bibinfo{author}{\bibfnamefont{R.}~\bibnamefont{Ryblewski}},
  \bibinfo{journal}{Phys. Rev. D} \textbf{\bibinfo{volume}{103}},
  \bibinfo{pages}{014030} (\bibinfo{year}{2021}{\natexlab{b}}),
  \eprint{2008.10976}.

\bibitem[{\citenamefont{Tinti and Florkowski}(2020)}]{Tinti:2020gyh}
\bibinfo{author}{\bibfnamefont{L.}~\bibnamefont{Tinti}} \bibnamefont{and}
  \bibinfo{author}{\bibfnamefont{W.}~\bibnamefont{Florkowski}}
  (\bibinfo{year}{2020}), \eprint{2007.04029}.

\bibitem[{\citenamefont{Son and Surowka}(2009)}]{Son:2009tf}
\bibinfo{author}{\bibfnamefont{D.~T.} \bibnamefont{Son}} \bibnamefont{and}
  \bibinfo{author}{\bibfnamefont{P.}~\bibnamefont{Surowka}},
  \bibinfo{journal}{Phys. Rev. Lett.} \textbf{\bibinfo{volume}{103}},
  \bibinfo{pages}{191601} (\bibinfo{year}{2009}), \eprint{0906.5044}.

\bibitem[{\citenamefont{Kharzeev and Son}(2011)}]{Kharzeev:2010gr}
\bibinfo{author}{\bibfnamefont{D.~E.} \bibnamefont{Kharzeev}} \bibnamefont{and}
  \bibinfo{author}{\bibfnamefont{D.~T.} \bibnamefont{Son}},
  \bibinfo{journal}{Phys. Rev. Lett.} \textbf{\bibinfo{volume}{106}},
  \bibinfo{pages}{062301} (\bibinfo{year}{2011}), \eprint{1010.0038}.

\bibitem[{\citenamefont{Montenegro
  et~al.}(2017{\natexlab{a}})\citenamefont{Montenegro, Tinti, and
  Torrieri}}]{Montenegro:2017rbu}
\bibinfo{author}{\bibfnamefont{D.}~\bibnamefont{Montenegro}},
  \bibinfo{author}{\bibfnamefont{L.}~\bibnamefont{Tinti}}, \bibnamefont{and}
  \bibinfo{author}{\bibfnamefont{G.}~\bibnamefont{Torrieri}},
  \bibinfo{journal}{Phys. Rev.} \textbf{\bibinfo{volume}{D96}},
  \bibinfo{pages}{056012} (\bibinfo{year}{2017}{\natexlab{a}}),
  \eprint{1701.08263}.

\bibitem[{\citenamefont{Montenegro
  et~al.}(2017{\natexlab{b}})\citenamefont{Montenegro, Tinti, and
  Torrieri}}]{Montenegro:2017lvf}
\bibinfo{author}{\bibfnamefont{D.}~\bibnamefont{Montenegro}},
  \bibinfo{author}{\bibfnamefont{L.}~\bibnamefont{Tinti}}, \bibnamefont{and}
  \bibinfo{author}{\bibfnamefont{G.}~\bibnamefont{Torrieri}},
  \bibinfo{journal}{Phys. Rev.} \textbf{\bibinfo{volume}{D96}},
  \bibinfo{pages}{076016} (\bibinfo{year}{2017}{\natexlab{b}}),
  \eprint{1703.03079}.

\bibitem[{\citenamefont{Gallegos et~al.}(2021)\citenamefont{Gallegos, G\"ursoy,
  and Yarom}}]{Gallegos:2021bzp}
\bibinfo{author}{\bibfnamefont{A.~D.} \bibnamefont{Gallegos}},
  \bibinfo{author}{\bibfnamefont{U.}~\bibnamefont{G\"ursoy}}, \bibnamefont{and}
  \bibinfo{author}{\bibfnamefont{A.}~\bibnamefont{Yarom}},
  \bibinfo{journal}{SciPost Phys.} \textbf{\bibinfo{volume}{11}},
  \bibinfo{pages}{041} (\bibinfo{year}{2021}), \eprint{2101.04759}.

\bibitem[{\citenamefont{Huang}(2021)}]{Huang:2020xyr}
\bibinfo{author}{\bibfnamefont{X.-G.} \bibnamefont{Huang}},
  \bibinfo{journal}{Nucl. Phys. A} \textbf{\bibinfo{volume}{1005}},
  \bibinfo{pages}{121752} (\bibinfo{year}{2021}), \eprint{2002.07549}.

\bibitem[{\citenamefont{Becattini and Lisa}(2020)}]{Becattini:2020ngo}
\bibinfo{author}{\bibfnamefont{F.}~\bibnamefont{Becattini}} \bibnamefont{and}
  \bibinfo{author}{\bibfnamefont{M.~A.} \bibnamefont{Lisa}},
  \bibinfo{journal}{Ann. Rev. Nucl. Part. Sci.} \textbf{\bibinfo{volume}{70}},
  \bibinfo{pages}{395} (\bibinfo{year}{2020}), \eprint{2003.03640}.

\bibitem[{\citenamefont{Florkowski
  et~al.}(2019{\natexlab{b}})\citenamefont{Florkowski, Kumar, and
  Ryblewski}}]{Florkowski:2018fap}
\bibinfo{author}{\bibfnamefont{W.}~\bibnamefont{Florkowski}},
  \bibinfo{author}{\bibfnamefont{A.}~\bibnamefont{Kumar}}, \bibnamefont{and}
  \bibinfo{author}{\bibfnamefont{R.}~\bibnamefont{Ryblewski}},
  \bibinfo{journal}{Prog. Part. Nucl. Phys.} \textbf{\bibinfo{volume}{108}},
  \bibinfo{pages}{103709} (\bibinfo{year}{2019}{\natexlab{b}}),
  \eprint{1811.04409}.

\bibitem[{\citenamefont{Speranza and Weickgenannt}(2021)}]{Speranza:2020ilk}
\bibinfo{author}{\bibfnamefont{E.}~\bibnamefont{Speranza}} \bibnamefont{and}
  \bibinfo{author}{\bibfnamefont{N.}~\bibnamefont{Weickgenannt}},
  \bibinfo{journal}{Eur. Phys. J. A} \textbf{\bibinfo{volume}{57}},
  \bibinfo{pages}{155} (\bibinfo{year}{2021}), \eprint{2007.00138}.

\bibitem[{\citenamefont{Bhadury
  et~al.}(2021{\natexlab{c}})\citenamefont{Bhadury, Bhatt, Jaiswal, and
  Kumar}}]{Bhadury:2021oat}
\bibinfo{author}{\bibfnamefont{S.}~\bibnamefont{Bhadury}},
  \bibinfo{author}{\bibfnamefont{J.}~\bibnamefont{Bhatt}},
  \bibinfo{author}{\bibfnamefont{A.}~\bibnamefont{Jaiswal}}, \bibnamefont{and}
  \bibinfo{author}{\bibfnamefont{A.}~\bibnamefont{Kumar}},
  \bibinfo{journal}{Eur. Phys. J. ST} \textbf{\bibinfo{volume}{230}},
  \bibinfo{pages}{655} (\bibinfo{year}{2021}{\natexlab{c}}),
  \eprint{2101.11964}.

\bibitem[{\citenamefont{Einstein and de~Haas}(1915)}]{dehaas:1915}
\bibinfo{author}{\bibfnamefont{A.}~\bibnamefont{Einstein}} \bibnamefont{and}
  \bibinfo{author}{\bibfnamefont{W.}~\bibnamefont{de~Haas}},
  \bibinfo{journal}{Deutsche Physikalische Gesellschaft, Verhandlungen}
  \textbf{\bibinfo{volume}{17}}, \bibinfo{pages}{152} (\bibinfo{year}{1915}).

\bibitem[{\citenamefont{Barnett}(1935)}]{Barnett:1935}
\bibinfo{author}{\bibfnamefont{S.~J.} \bibnamefont{Barnett}},
  \bibinfo{journal}{Rev. Mod. Phys.} \textbf{\bibinfo{volume}{7}},
  \bibinfo{pages}{129} (\bibinfo{year}{1935}),
  \urlprefix\url{https://link.aps.org/doi/10.1103/RevModPhys.7.129}.

\bibitem[{\citenamefont{Siddique et~al.}(2019)\citenamefont{Siddique, Liang,
  Lisa, Wang, and Xu}}]{Siddique:2017ddr}
\bibinfo{author}{\bibfnamefont{I.}~\bibnamefont{Siddique}},
  \bibinfo{author}{\bibfnamefont{Z.-T.} \bibnamefont{Liang}},
  \bibinfo{author}{\bibfnamefont{M.~A.} \bibnamefont{Lisa}},
  \bibinfo{author}{\bibfnamefont{Q.}~\bibnamefont{Wang}}, \bibnamefont{and}
  \bibinfo{author}{\bibfnamefont{Z.-B.} \bibnamefont{Xu}},
  \bibinfo{journal}{Chin. Phys. C} \textbf{\bibinfo{volume}{43}},
  \bibinfo{pages}{014103} (\bibinfo{year}{2019}), \eprint{1710.00134}.

\bibitem[{\citenamefont{Poskanzer and Voloshin}(1998)}]{Poskanzer:1998yz}
\bibinfo{author}{\bibfnamefont{A.~M.} \bibnamefont{Poskanzer}}
  \bibnamefont{and} \bibinfo{author}{\bibfnamefont{S.~A.}
  \bibnamefont{Voloshin}}, \bibinfo{journal}{Phys. Rev. C}
  \textbf{\bibinfo{volume}{58}}, \bibinfo{pages}{1671} (\bibinfo{year}{1998}),
  \eprint{nucl-ex/9805001}.

\bibitem[{\citenamefont{Jackson}(1998)}]{Jackson:1998nia}
\bibinfo{author}{\bibfnamefont{J.~D.} \bibnamefont{Jackson}},
  \emph{\bibinfo{title}{{Classical Electrodynamics}}}
  (\bibinfo{publisher}{Wiley}, \bibinfo{year}{1998}), ISBN
  \bibinfo{isbn}{978-0-471-30932-1}.

\bibitem[{\citenamefont{Leader}(2011)}]{Leader:2001gr}
\bibinfo{author}{\bibfnamefont{E.}~\bibnamefont{Leader}},
  \emph{\bibinfo{title}{{Spin in particle physics}}}, vol.~\bibinfo{volume}{15}
  (\bibinfo{year}{2011}), ISBN \bibinfo{isbn}{978-0-511-87418-5,
  978-0-521-35281-9, 978-0-521-02077-0}.

\bibitem[{\citenamefont{Itzykson and Zuber}(1980)}]{Itzykson:1980rh}
\bibinfo{author}{\bibfnamefont{C.}~\bibnamefont{Itzykson}} \bibnamefont{and}
  \bibinfo{author}{\bibfnamefont{J.~B.} \bibnamefont{Zuber}},
  \emph{\bibinfo{title}{{Quantum Field Theory}}}, International Series In Pure
  and Applied Physics (\bibinfo{publisher}{McGraw-Hill}, \bibinfo{address}{New
  York}, \bibinfo{year}{1980}), ISBN \bibinfo{isbn}{9780486445687, 0486445682},
  \urlprefix\url{http://dx.doi.org/10.1063/1.2916419}.

\end{thebibliography}

%\begin{thebibliography}{99}
%\end{thebibliography}

%%%%%%%%%%%%%%%%%%%%%%%%%%%%%%%%%%%%%%%%%%
\end{document}